\documentclass[sigconf]{acmart}

\renewcommand\footnotetextcopyrightpermission[1]{} 
\usepackage{booktabs} 
\usepackage{enumerate}
\usepackage{epstopdf}
\usepackage{epsfig}
\usepackage{subfigure}

\newcommand{\tabincell}[2]{\begin{tabular}{@{}#1@{}}#2\end{tabular}}

\newcommand{\hank}{\textcolor{black}}
\newcommand{\ruby}{\textcolor{black}}

\begin{document}
\title{Secure Pick Up: Implicit Authentication When You Start Using the Smartphone}

\author{Wei-Han Lee}
\affiliation{%
  \institution{Princeton University}
}
\email{weihanl@princeton.edu}

\author{Xiaochen Liu}
\affiliation{%
  \institution{University of Southern California}
}
\email{liu851@usc.edu}

\author{Yilin Shen}
\affiliation{%
  \institution{Samsung Research America}
}
\email{yilin.shen@samsung.com}

\author{Hongxia Jin}
\affiliation{
  \institution{Samsung Research America}
}
\email{hongxia.jin@samsung.com}

\author{Ruby B. Lee}
\affiliation{%
  \institution{Princeton University}
}
\email{rblee@princeton.edu}

%
%


\begin{abstract}
We propose Secure Pick Up (SPU), a convenient, lightweight, in-device, non-intrusive and automatic-learning system for smartphone user authentication. Operating in the background, our system implicitly observes users' phone pick-up movements, the way they bend their arms when they pick up a smartphone to interact with the device, to authenticate the users. 

Our SPU outperforms the state-of-the-art implicit authentication mechanisms in three main aspects: 1) SPU automatically learns the user's behavioral pattern without requiring a large amount of training data (especially those of other users) as previous methods did, making it more deployable. Towards this end, we propose a weighted multi-dimensional Dynamic Time Warping (DTW) algorithm to effectively quantify similarities between users' pick-up movements; 2) SPU does not rely on a remote server for providing further computational power, making SPU efficient and usable even without network access; and 3) our system can adaptively update a user's authentication model to accommodate user's behavioral drift over time with negligible overhead.

Through extensive experiments on real world datasets, we demonstrate that SPU can achieve authentication accuracy up to $96.3\%$ with a very low latency of $2.4$ milliseconds. It reduces the number of times a user has to do explicit authentication by $32.9\%$, while effectively defending against various attacks.
\end{abstract}

\keywords{Authentication; Security; Privacy; Machine Learning; Smartphone; Dynamic Time Warping; Mobile System}

\maketitle
\section{Introduction}
Mobile devices such as smartphones and tablets are rapidly becoming our means for entering the Internet and online social networks. They also store sensitive and personal information, such as email addresses or bank account information of users. The hardware of today's mobile devices is quite capable with multi-core gigahertz processors, and gigabytes of memory and solid-state storage. Their relatively low cost, ease of use and `always on' connectivity provide a suitable platform for many day-to-day tasks involving financial transactions and sensitive data, making mobile devices attractive attack targets (e.g., see attacks against the Apple iOS and Google Android platforms in \cite{nachenberg2011window}).

Passwords are currently one of the most common forms for user authentication in mobile devices. However, they suffer from several weaknesses. Passwords are vulnerable to guessing attacks \cite{weir2009sp,kelley2012sp,bonneau2012sp,uellenbeck2013CCS,ma2014sp} or password reuse \cite{das2014ndss}. The usability issue is also a serious factor, since users do not like to have to enter, and reenter, passwords \cite{report_convenience, shi2011implicit}. A recent study in \cite{consumer} shows that $64\%$ of users do not use passwords or PINs as an authentication mechanism on their smartphones.

Recently, more and more smartphones are equipped with fingerprint scanners, making authentication through fingerprints quite popular. However, such mechanisms also suffer from several weaknesses. It is possible to trick the scanner by using a gelatin print mold over a real finger. In addition, the response time for the fingerprint scanner to unlock the smartphone is often more than one second \cite{report_fingerprint}, degrading the usability of fingerprint-based authentication.

Other biometric-based authentication mechanisms (e.g., via face and keystroke dynamics) are also unreliable and vulnerable to forgery attacks \cite{trewin2012acsac,tey2013ndss}. For instance, an attacker can obtain a photo of the targeted user (e.g., via Facebook) and present it in front of the camera to spoof face recognition on smartphones. Furthermore, these authentication mechanisms require frequent user participation, hindering their deployment in real world scenarios. Hence, it is important to design secure and convenient authentication methods for smartphone users, the topic of this paper. 

Behavior-based authentication mechanisms are recently proposed to implement convenient and implicit authentication which does not require frequent user participation and can reduce the user's efforts (e.g., the number of times) needed to unlock their smartphones. Behavior-based authentication is increasingly gaining popularity since mobile devices are often equipped with sensors such as accelerometer, gyroscope, magnetometer, camera, microphone, GPS and so on. Implicit authentication relies on a distinguishable behavioral pattern of the user, which is accomplished by building the users' profiles~\cite{cc5,cc7,cc2,cc4,cc6,lee2015icissp,frank2013touchalytics,shi2011implicit, lin2012new,eagle2009eigenbehaviors, riva2012progressive,conti2011swing,lee2016implicit,lee2015implicit,lee2017implicit}. If a newly-detected user behavior is consistent with the behavior profile stored in the smartphone, the device will have high confidence that no explicit authentication action is required. Otherwise, if the newly-detected behavior deviates significantly from the stored behavior profile, alternative explicit authentication mechanisms should be triggered, such as requiring the user to enter a password, PIN or checking his/her fingerprint. 

Existing behavior-based authentication systems exploit machine learning techniques to achieve good security performance~\cite{cc5,cc2,cc4,cc6,lee2015icissp,frank2013touchalytics}. However, these systems have several limitations for real world user authentication: 1) they need a large amount of training data (including other users' data) to learn an authentication classifier, which may violate users' privacy and thus hinder users' motivation to utilize these systems; 2) their training process is usually computationally complicated, which requires additional computational services, e.g., cloud computing, thus requiring users to trust the remote server and always have network connection; 3) their system updating process for capturing the user's behavioral drift over time is also quite complex. 

Other behavior-based authentication mechanisms exploit specific contexts of users' behavior, e.g., how do users walk~\cite{cc7}, and how do users answer a phone call~\cite{conti2011swing}, for authentication. However, their corresponding experiments require users to follow restricted patterns for authentication, e.g., walk straight ahead at the same speed~\cite{cc7} or answer a call when the phone is on a table in front of a user~\cite{conti2011swing}. These constraints are unrealistic for extracting effective behavior patterns of users, making these systems impractical for real world authentication.

To address these issues, we propose a lightweight, in-device, non-intrusive and automatic-learning authentication system, called Secure Pick Up (SPU), which can be broadly deployed in real world mobile devices. Our system aims to utilize a simple and general behavioral pattern of smartphone users, the way people bend their arms when they pick up a phone to interact with the device, to implicitly authenticate the users. For a smartphone that installs our SPU application, the device starts extracting a user's pick-up pattern from his/her arm movements when picking up a phone, and then the system determines whether the current user is legitimate or not. If the user's current behavior conforms to the established behavior profile stored in the smartphone, the user passes the authentication and can have access to the smartphone. If the user's current behavior deviates from the established behavior profile, the device would present explicit authentication challenges, e.g., input of a password, PIN or fingerprint. If these backup explicit authentication mechanisms pass, the user is allowed access to the smartphone and the user's profile stored in the smartphone is updated consequently; otherwise, the user is denied access. This paper aims to answer the question of whether we could build and deploy such a model in a practical, convenient and secure manner on today's mobile devices. Our key contributions include:

\begin{enumerate}[$\bullet$]
	\item{We design a behavior-based implicit authentication system, SPU, by exploiting users' behavioral patterns recorded by smartphone sensors when they bend their arms to pick up a phone. SPU can automatically learn a user's behavioral pattern in an accurate, efficient and stealthy manner. Furthermore, SPU does not require a large amount of training data of other users as previous work did, making our system easier to deploy in real world applications.}
	\item {Our system (including the profile updating process) can be implemented efficiently and entirely on personal smartphones. It does not require any additional computational services, e.g., cloud computing. To the best of our knowledge, it is the first using only a device's resources for implicit authentication, making SPU efficient and usable even without network access. For instance, our system can adaptively update the user's authentication model over time with rather low overhead, consuming negligible power of $2\%$.} 
 \item{We propose an effective Dynamic Time Warping (DTW) algorithm to quantify similarities between users' pick-up patterns. More specifically, we modify the traditional DTW algorithm and propose a weighted multi-dimensional DTW technique to accommodate the multiple dimensions of sensor data in our setting, and to further improve authentication performance. Extensive experimental results verify the effectiveness of our method which can achieve high accuracy up to $96.3\%$ in $2.4$ milliseconds. Furthermore, we demonstrate that SPU can reduce a user's efforts by $32.9\%$ to unlock his/her smartphone providing a more user-friendly experience and encouraging more users to protect access to their devices.} 
\item{Finally, our system is robust to various types of attackers, including the serious ones that observe victims' behaviors many times. For instance, our SPU can achieve $0\%$ false acceptance rate (FAR) and $18\%$ false rejection rate (FRR) for authenticating smartphone users under the worst case mimicry attacks (educated attacks).}
\end{enumerate}

\section{System Design}\label{sec:system}
The main objective of our SPU system is to increase the convenience for smartphone users by reducing their efforts (e.g., the number of times) to unlock the smartphone while guaranteeing their security through preventing unauthorized access to the smartphone. We now describe the threat model, design goals, key ideas and system architecture for SPU.

\subsection{Threat Model}\label{threat}
Compared to personal computers, smartphones are more easily lost or stolen, giving attackers more opportunity to obtain the sensitive data stored in the smartphones. We assume that the attackers have physical access to the smartphone and can even monitor and mimic the user's pick-up behavior. Therefore, they can launch mimicry attacks, to impersonate the legitimate user's behavior.
Specifically, we consider three different levels of attacks as follows.

\begin{enumerate}[$\bullet$]
\item Random Attack (RA): With no prior knowledge of the user's pick-up behavior, a RA attacker randomly picks up the smartphone and wishes to pass the authentication system. This is equivalent to a brute force attack against text-based password schemes.
\item Context-Aware Attack (CAA): In a context-aware attack, an adversary knows the place where the user picks up his/her smartphone, but has not observed how the user does it.
\item Educated Attack (EA): In an educated attack, an adversary has observed how and where the user picks up his/her smartphone.
\end{enumerate}

In our SPU system, we consider a single-user model, which is in line with current smartphone usage scenarios. For multi-user models, our system can be generalized in a straightforward manner to incorporate multiple profiles (e.g., family members, guests) for progressive authentication as discussed in \cite{liu2009xshare, ni2009diffuser}. Furthermore, we assume the availability of low-cost sensors in mobile devices for detecting a user's presence and behavior. Indeed, the sensors used in our implementation are the accelerometer and gyroscope, which are widely available in today's mobile devices. As more sensors become pervasive, they can easily be folded into our system.

\subsection{Design Goals}\label{design_goal}
Our system is designed to increase the convenience of smartphone users while guaranteeing their security, through implicitly authenticating the users in an unobtrusive manner. Furthermore, the whole authentication process should be implemented stealthily and efficiently. Overall, our design goals for the SPU system are:

\begin{enumerate}[$\bullet$]
\item Accurate: the authentication system should not incorrectly authenticate a user.

\item Rapid Enrollment and Updating: creating new user accounts or updating pick-up profiles for existing users should be quick.

\item Rapid Authentication: the response time for the authentication system must be short, for the system to be usable in reality.
\item Implicit: the authentication system should neither interrupt user-smartphone interactions nor need explicit user participation during the authentication process.

\item Unobtrusive: the authentication system should be completely unobtrusive and should not invade the user's privacy; the user should be comfortable when using our system.

\item Light-weight: the authentication system should not require intensive computations.

\item Device only: the authentication system should work efficiently and entirely on mobile devices only even without network access. It should not depend on auxiliary training data of other users or additional computational capabilities, e.g., cloud computing.
\end{enumerate}

\subsection{Key Ideas}\label{sec:key}
Our SPU system is designed to achieve all the design goals in Section~\ref{design_goal}. To increase the convenience for users and detect unauthorized access to the smartphone as soon as possible, it is required that we authenticate the users when they start using the smartphone. Therefore, we consider using the users' arm movements 
when they pick up their smartphones as a distinguishable behavior to authenticate the users. Our key idea stems from the observation that users' behavioral patterns are different from person to person when they start using their smartphones, from the time they pick up the phone to the time they press the \emph{home} button or \emph{power} button. More specifically, we extract the `pick-up signal' from the user's arm movements measured by sensors (accelerometer and gyroscope) embedded in the smartphone. 

\begin{figure}
\centering
\epsfig{file=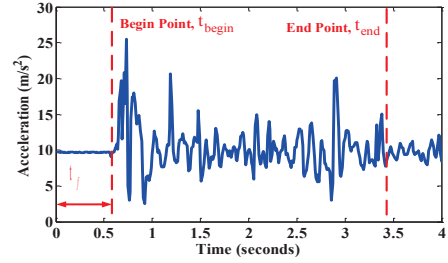, height=1.5 in, width=2.7 in}
\caption{A real world instance of a user's pick-up movement. When a \emph{wake up} signal is detected (\emph{home} button or \emph{power} button is pressed in the sleep mode) corresponding to the {end point} $t_{end}$, we backtrack the sensor measurements to find the {begin point} $t_{begin}$ after detecting a flat signal lasting a period of $t_f$.}
\label{fig:signal}
\end{figure}

To extract users' pick-up movements, we first define a particular user action and call it a `trigger-action'. Here, we utilize the `wake up' signal of a smartphone such as pressing the \emph{home} button or \emph{power} button in the sleep mode, as the trigger action \footnote{\hank{In our experiments, we used the \emph{home} button or \emph{power} button as the `trigger-action'. Our method can be easily integrated with new trigger-actions, e.g., the automatic wake-up feature in \ruby{the} iphone 7.}}. Whenever a trigger-action is performed, we extract the pick-up signal from the measurements of the accelerometer and gyroscope (described below). That is to say, our system authenticates the user only when the smartphone is triggered to wake up from the sleep mode. Note that there is no necessity to authenticate the user when the smartphone is locked. 
 
Figure \ref{fig:signal} shows a real world instance for the extracted signal stream that describes a user's pick-up movements from measurements collected by the \emph{accelerometer}. When our system detects the \emph{home} button signal or \emph{power} button signal during the \emph{sleep} mode, we record the time as the end of the pick-up signal $t_{end}$, and back-track the accelerometer measurements to construct the pick-up signal. If we detect a flat signal lasting for a time period of $t_f$, we consider the end time of the flat signal as the beginning of the pick-up signal $t_{begin}$ as shown in Figure \ref{fig:signal}. 

In order to backtrack the pick-up signal, we need to record the entire time-series measurements of the accelerometer and gyroscope, while the smartphone is in the \emph{sleep} mode. In Section \ref{sec:overhead}, we will show that this sensor measurement process is efficient, only costing an additional $2\%$ in power consumption of the smartphone.

Note that we only consider \ruby{authenticating pick-up movements from a stable state in our SPU system. We will show in Section~\ref{second_exp} that this type of pick-up movement (from a stable state) constitutes the most important pick-up characteristic of users.}

After extracting the pick-up signal, we propose a weighted multi-dimensional Dynamic Time Warping algorithm to effectively quantify similarities between users' pick up movements for authentication (detailed process will be discussed in Section~\ref{sec:newDTW}). More specifically, we modify the traditional DTW algorithm to accommodate the multi-dimensional sensor data in our setting, to further improve authentication performance. 

We will show the distinguishable properties of users' pick-up patterns in Section \ref{sec:experiments}. We will show that the pick-up signals are still distinguishable even under impersonation attacks in Section \ref{sec:security}. Furthermore, our SPU system can significantly reduce users' efforts to unlock their smartphones as will be discussed in Section~\ref{diff_sensor}. 

Unlike previous work, our SPU does not require a large amount of training data for learning a complex authentication classifier, and any additional computational capability of cloud servers, therefore more users would be motivated to use our system. In addition, our system can be easily combined with the state-of-the-art re-authentication systems \cite{cc5,cc6,lee2016implicit,lee2017implicit} to further improve the security of the smartphone.

\begin{figure}
\centering
\epsfig{file=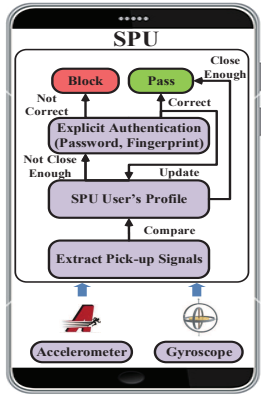, height=2.2 in, width=1.5 in}
\caption{The flowchart of our SPU system.}
\label{fig:mechanism}
\end{figure}

\subsection{System Architecture}\label{sec:mechanism}
Our system is designed for today's smartphones which are equipped with rich sensing capabilities. It could also be generally applied to tablets and other types of wearable devices such as smartwatches. Figure \ref{fig:mechanism} shows the flowchart of our SPU system. System operation is in four phases:

{\bf{Enrollment}}: When a user first enrolls in our SPU system, he/she is asked to pick up his/her smartphone in the same way as in his/her normal life. Our system then establishes the user's pick-up profile by extracting the pick-up signal and storing it in the smartphone.

{\bf{Extracting pick-up signals}}: Our system keeps monitoring and recording the measurements of the accelerometer and gyroscope when the smartphone is in the sleep mode until it is picked up. We extract the pick-up signals from these sensor measurements in the enrollment phase and afterwards (detailed process discussed in Section~\ref{sec:key}).

{\bf{Authentication}}: After extracting the pick-up signal, we compare the new incoming measurements (signal) with the user's pick-up profile stored in the smartphone by utilizing our proposed weighted multi-dimensional DTW technique (will be discussed in Section~\ref{sec:newDTW}). 

{\bf{Post-Authentication}}: If the pick-up signal is authenticated as coming from the legitimate user, this testing passes and the current user can access the information and resources in the smartphone. Otherwise, the smartphone would request an explicit authentication, e.g., password, PIN or fingerprint, from the current user. We emphasize, however, that the desired response to such situations is a matter of policy. Furthermore, the stored user's profile will be updated to accommodate the user's behavioral drift if the correct explicit authentication is provided. Otherwise, no access to the smartphone is allowed.

\section{Data collection} \label{sec:dataset}
\subsection{Sensor Selection}
There are various built-in sensors in today's smartphones, from which we aim to choose a small set of sensors that can accurately represent a user's pick-up behavioral pattern. In this paper, we consider the following two sensors that are commonly embeded in current smartphones: the accelerometer and the gyroscope~\cite{google}. 

These two sensors represent different levels of information about the user's behavior, and are often called a 6-axis motion detector. The accelerometer records larger motion patterns of users such as how they move their arms or walk \cite{cc7}, while the gyroscope records fine-grained motions of users such as how they hold the smartphone \cite{cc8}. Furthermore, these sensors do not require the user's permission when requested by mobile applications \cite{google_permission}, making them useful for background monitoring as in our implicit authentication systems.

\subsection{Dataset Collection}\label{sec:data}
We utilize the open-source Android system as our implementation platform. We develop an Android application to implement SPU on Andriod smartphones. Note that our methods are not limited to this platform and can be easily applied to other platforms such as the Apple iOS platform on an iPhone. 

In our experiments, each data sample is a time-series measurement collected by the accelerometer and gyroscope, which captures the user's behavioral pattern when picking up the smartphone.
In our user study, we consider three experimental scenarios and describe the detailed settings for each experiment as follows.
All the participants were shown the app that is installed in their phones. All of the participants volunteered to participate in our experiments. There is no security breach on users' data in smartphones since we collect data and do the authentication attempts offline.

The first experiment was conducted under a lab setting, aiming to provide fundamental intuition for our SPU system. We collected sensor data from $24$ users whose detailed demographics are described in Section~\ref{first_exp}. We asked each user to pick up the smartphone in $6$ different places while sitting or standing \footnote{2 places are at a user's right hand side, another 2 places are in front of the user, and another 2 places are at a user's left hand side. In each of these three directions, one place is close while the other place is far.}. For each scenario, we collected $10$ samples of the pick-up movement for each user, under the $12$ situations ($6$ places $\times$ $2$ user states). Therefore, we collected $2,880$ (i.e., $24\times 12\times 10$) pick-up samples in total. We will describe the detailed analysis for the first experiment in Section~\ref{first_exp}.

The second experiment was conducted under a more realistic setting which is designed to verify the effectiveness of our SPU system in real world applications. The same $24$ users were invited to install our application on their own smartphones and use them freely in their normal lives for \hank{a week.} From the collected data, we extracted $3,115$ pick-up movement samples for these users. We will analyze the overall authentication performance of our system in real world scenarios in Section~\ref{second_exp}.

Our third experiment was designed to analyze the security performance of SPU in defending against multiple attacks (e.g., impersonation attacks) as discussed in Section~\ref{threat}. In this experiment, we randomly select $6$ out of the $24$ users as victims and randomly select $12$ out of the other $18$ users (different from the victims) as adversaries. The experimental setting is the same as the first experiment. The only difference is that the adversaries are trying to mimic the victims under different levels of prior knowledge. Specifically, these adversaries perform the three attacks in Section~\ref{threat} respectively, and the detailed attack processes are described as follows:

\begin{enumerate}[$\bullet$]
\item Random Attack (RA): The random attacker tries to use the victim's smartphone without knowing any information about the victim. In total, we collected $12 \times 6 \times 10 = 720$ samples\footnote{In our experiments, we considered $12$ attackers, $6$ victims and $10$ repeated iterations for each user's pick-up movement.} of the pick-up signals under the random attack.

\item Context-Aware Attack (CAA): We provided a context-aware attacker who is informed of the place where the victim picked up the smartphone. Note that these attackers have not observed how the victim picked up the smartphone. We also collected $720$ pick-up samples under the context-aware attack.

\item Educated Attack (EA):  The victim user's behavior was recorded by a VCR and is clearly visible to the attacker. The attacker was asked to watch the video and mimic the victim's behavior to the best of his/her ability. In total, we also collected $720$ pick-up samples under the educated attack.
\end{enumerate}

We will discuss the security analysis for the third experiment in Section~\ref{sec:security}.

\section{SPU Authentication Algorithms}
We now describe the design of our authentication algorithm which aims to achieve the design goals in Section~\ref{design_goal}. 

Previous implicit authentication algorithms exploit machine learning techniques to achieve good authentication performance \cite{cc5,cc7,cc2,cc4,cc6,lee2015icissp,frank2013touchalytics}. However, we identify characteristics that the smartphone implicit authentication exhibits that are not well aligned with the requirements of machine-learning techniques. These include: 1) lack of training data especially those of other users; 2) fundamental limitations in computation capabilities for the training process and the updating process.

To overcome these challenges, we aim to design an implicit, lightweight and in-device authentication algorithm by matching the new incoming pick-up signal with the pick-up profile stored in the smartphone, instead of the complicated machine learning techniques of previous methods. Furthermore, the time duration of a pick-up movement varies across time and across users, \ruby{and} typically is within the range of $0.5$ to $4$ seconds. Therefore, our matching process should also automatically cope with time deformations and different speeds associated with time-dependent sensor data. 

Towards these goals, we consider using the dynamic time warping technique \cite{muller2007dynamic} to carefully measure the distance between two time-series sensor data which may vary in time or speed. In DTW, the sequences are warped in a nonlinear fashion to match each other. \ruby{It} has been successfully applied to compare different speech patterns in automatic speech recognition and other applications in the data mining community. Furthermore, we propose an effective weighted multi-dimensional DTW to accommodate our setting where the collected sensor data are of multiple dimensions, thus taking the different distinguishing power of each sensor dimension into consideration. 

\subsection{Data Pre-processing}
Our system keeps monitoring and collecting the measurements of the accelerometer and gyroscope in the background, while the smartphone is in \emph{sleep} mode. When the \emph{wake up} signal (e.g., \emph{home} button or \emph{power} button is pressed in the sleep mode) is detected, our SPU records the time as the ending of the pick-up signal and back-tracks the collected data to find the beginning of the pick-up signal, as described earlier in Section \ref{sec:key}.

\subsection{DTW-based Authentication Algorithm}

\subsubsection{One-Dimensional DTW}\label{one_DTW}
DTW is a well-known technique to find the optimal alignment between two given (time-dependent) sequences $X :=(x_1, x_2,\dots ,x_N )$ of length $N \in \mathbb{N}$ and $Y := (y_1, y_2,\dots ,y_M)$ of length $M \in \mathbb{N}$ under certain restrictions. While there is a surfeit of possible distance measures for time-series data, empirical evidence has shown that DTW is exceptionally difficult to beat. Ding et al. in \cite{ding2008querying} tested the most cited distance measures on $47$ different datasets, and no method consistently outperforms DTW. Therefore, in our system, we utilize DTW to measure the distance between users' pick-up signals. 

DTW calculates the distance of two sequences using dynamic programming \cite{bertsekas1996dynamic}. It constructs an $N$-by-$M$ matrix, where the $(i,j)$-th element is the minimum distance (called local distance) between the two sequences that end at points $x_i$ and $y_j$ respectively. 
An $(N,M)$-warping path $p =(p_1,p_2,\cdots,p_L)$ is a contiguous set of matrix elements which defines an alignment between two sequences $X$ and $Y$ by aligning the element $x_{n_l}$ of $X$ to the element $y_{m_l}$ of $Y$. The boundary condition enforces that the first elements of $X$ and $Y$ as well as the last elements of $X$ and $Y$ are aligned to each other. The total distance $d_p(X, Y)$ of a warping path $p$ between $X$ and $Y$ with respect to the local distance measure $d$ is defined as $d_p(X,Y) = \sum_{l=1}^L d(x_{n_l},y_{m_l})$. Therefore, the DTW for one dimensional time-series data can be computed as 
\begin{equation}\label{eq_1}
DTW_1(X,Y) = \min d_p(X,Y)
\end{equation}
\begin{figure*}
\centering
\subfigure[Accelerometer $x$:original signal]{
\label{fig:acc1_original}
\epsfig{file=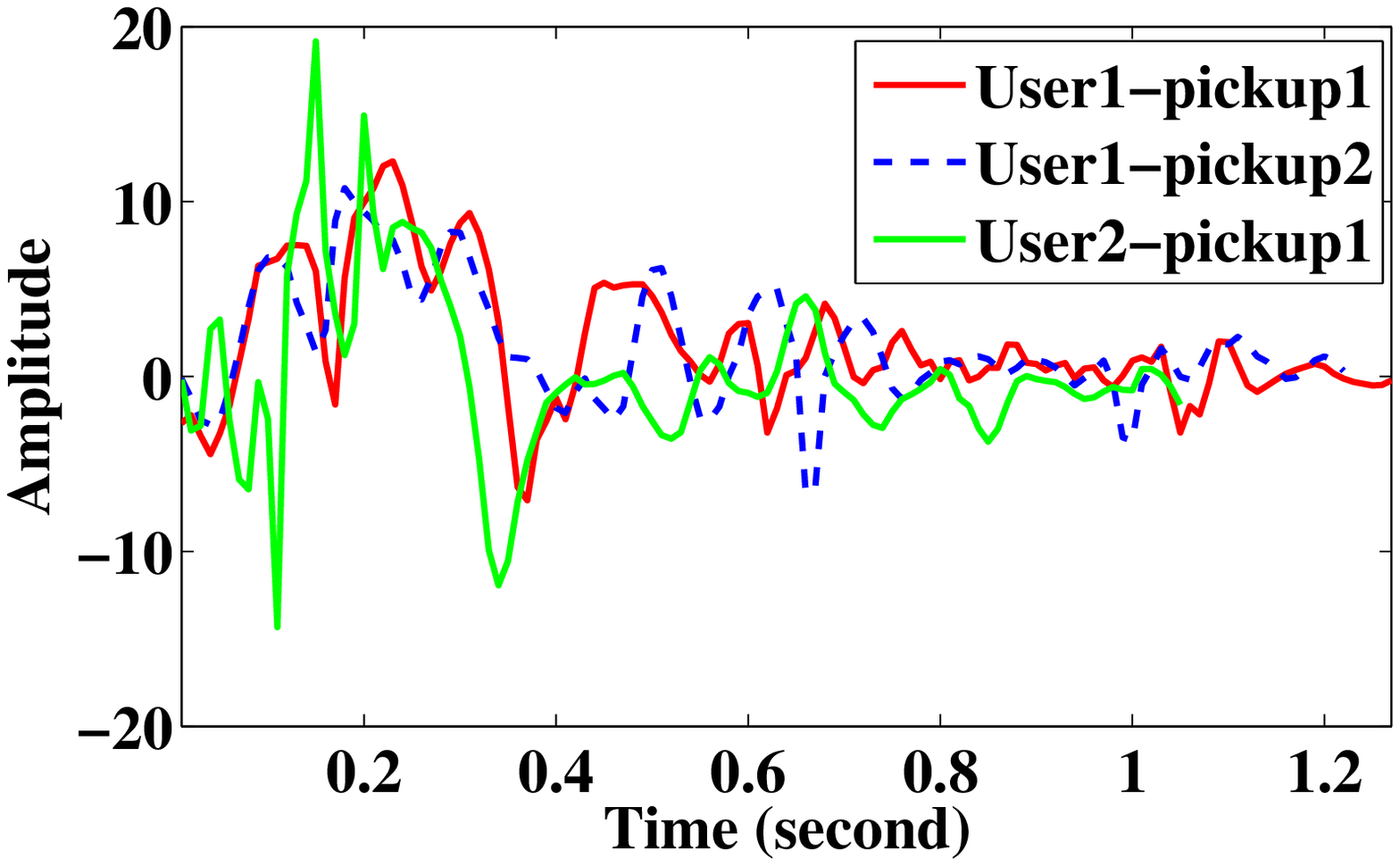, height=1.5 in, width=2.1 in}}
\subfigure[Accelerometer $y$:original signal]{
\label{fig:acc2_orignal}
\epsfig{file=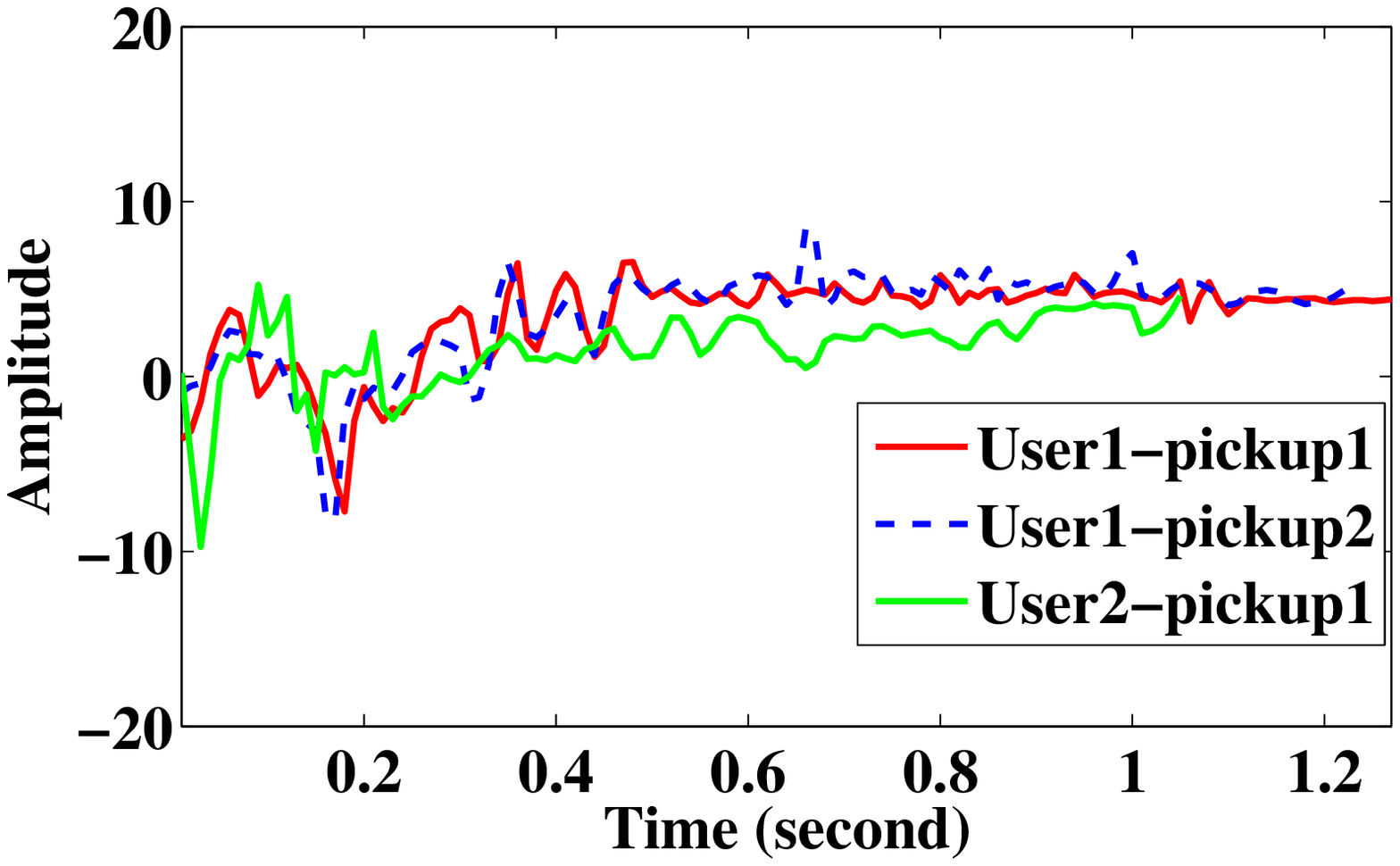, height=1.5 in, width=2.1 in}}
\subfigure[Accelerometer $z$:original signal]{
\label{fig:acc3_orignal}
\epsfig{file=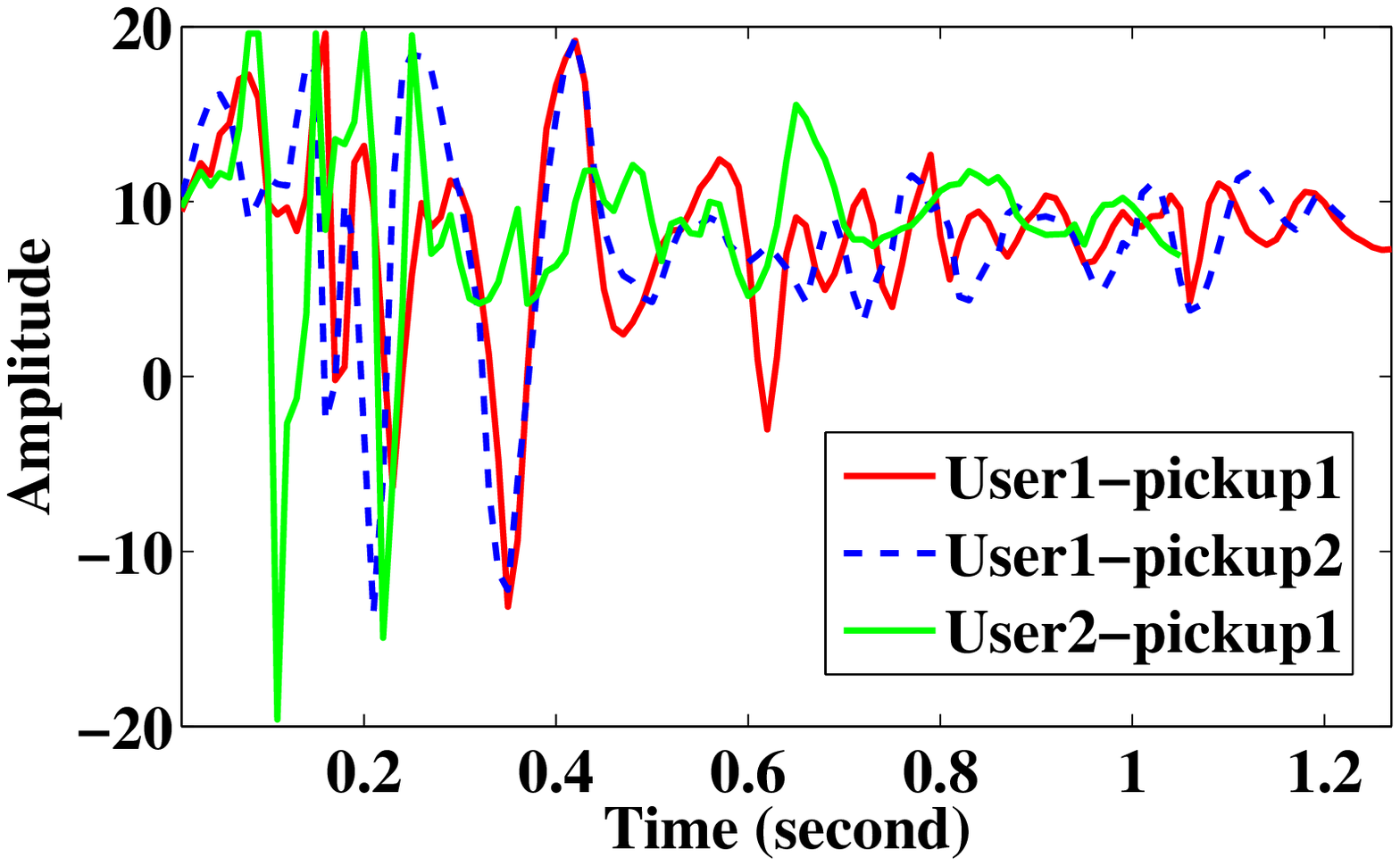, height=1.5 in, width=2.1 in}}
\subfigure[Accelerometer $x$:signal after DTW]{
\label{fig:acc1_dtw}
\epsfig{file=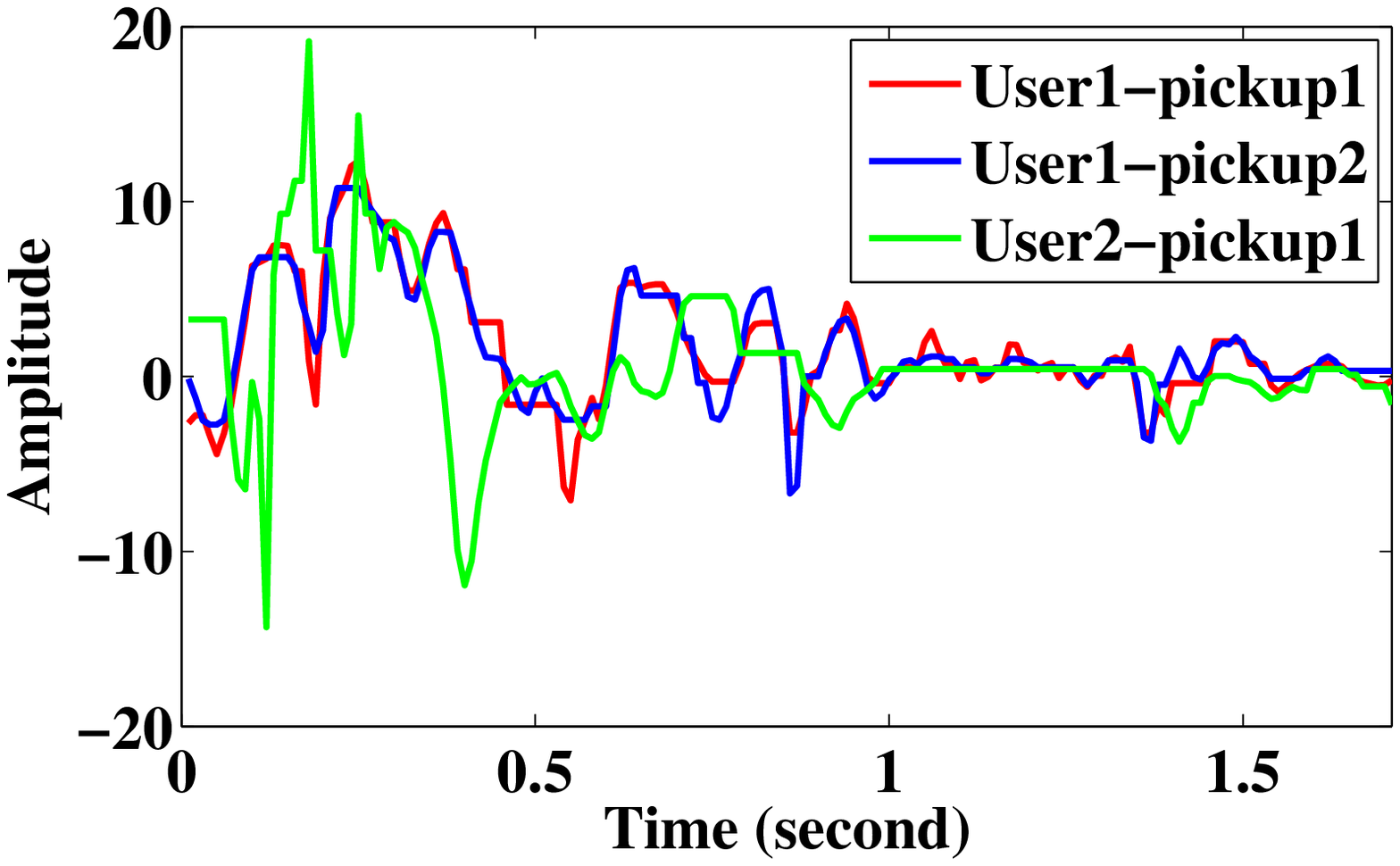, height=1.5 in, width=2.1 in}}
\subfigure[Accelerometer $y$:signal after DTW]{
\label{fig:acc2_dtw}
\epsfig{file=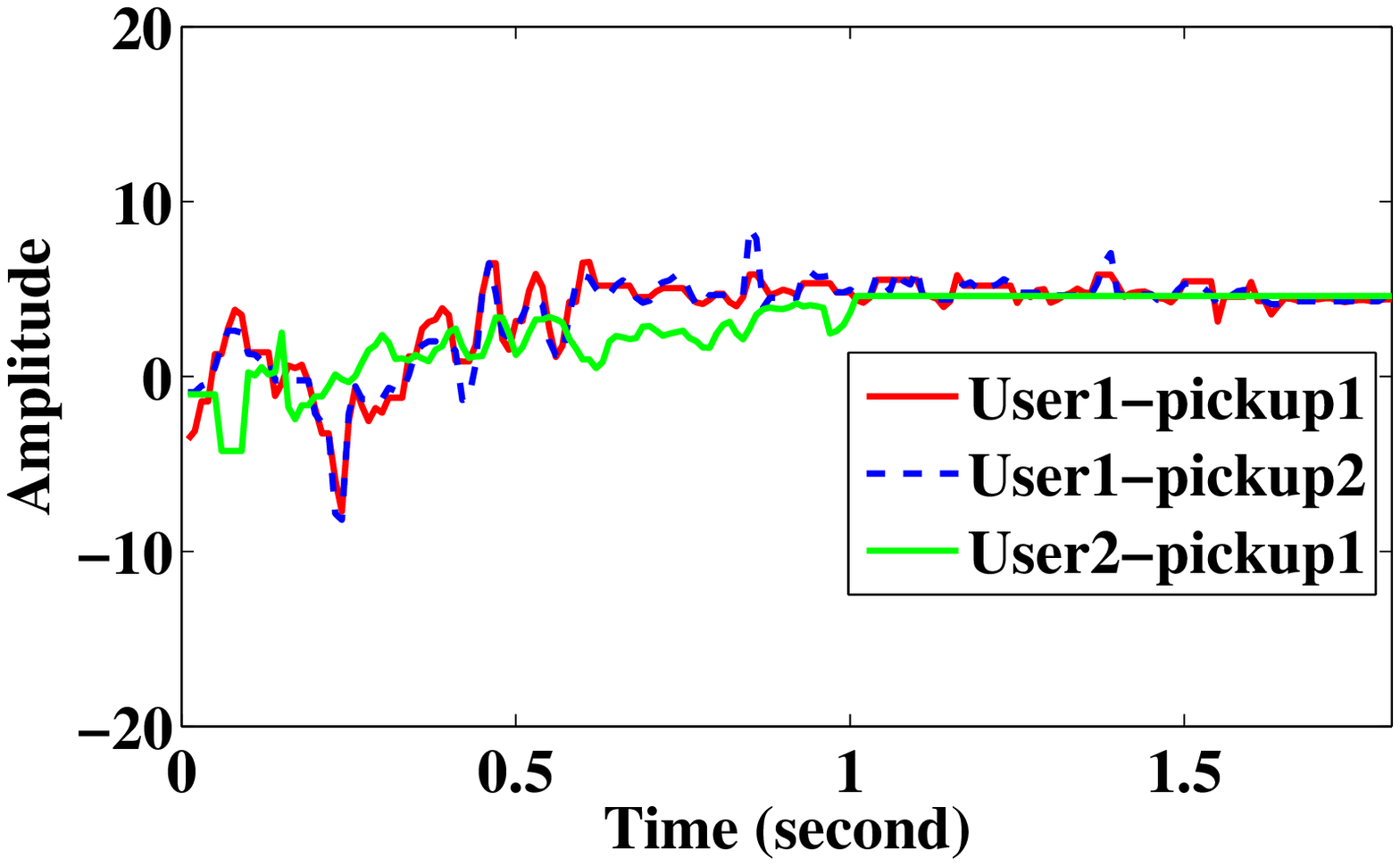, height=1.5 in, width=2.1 in}}
\subfigure[Accelerometer $z$:signal after DTW]{
\label{fig:acc3_dtw}
\epsfig{file=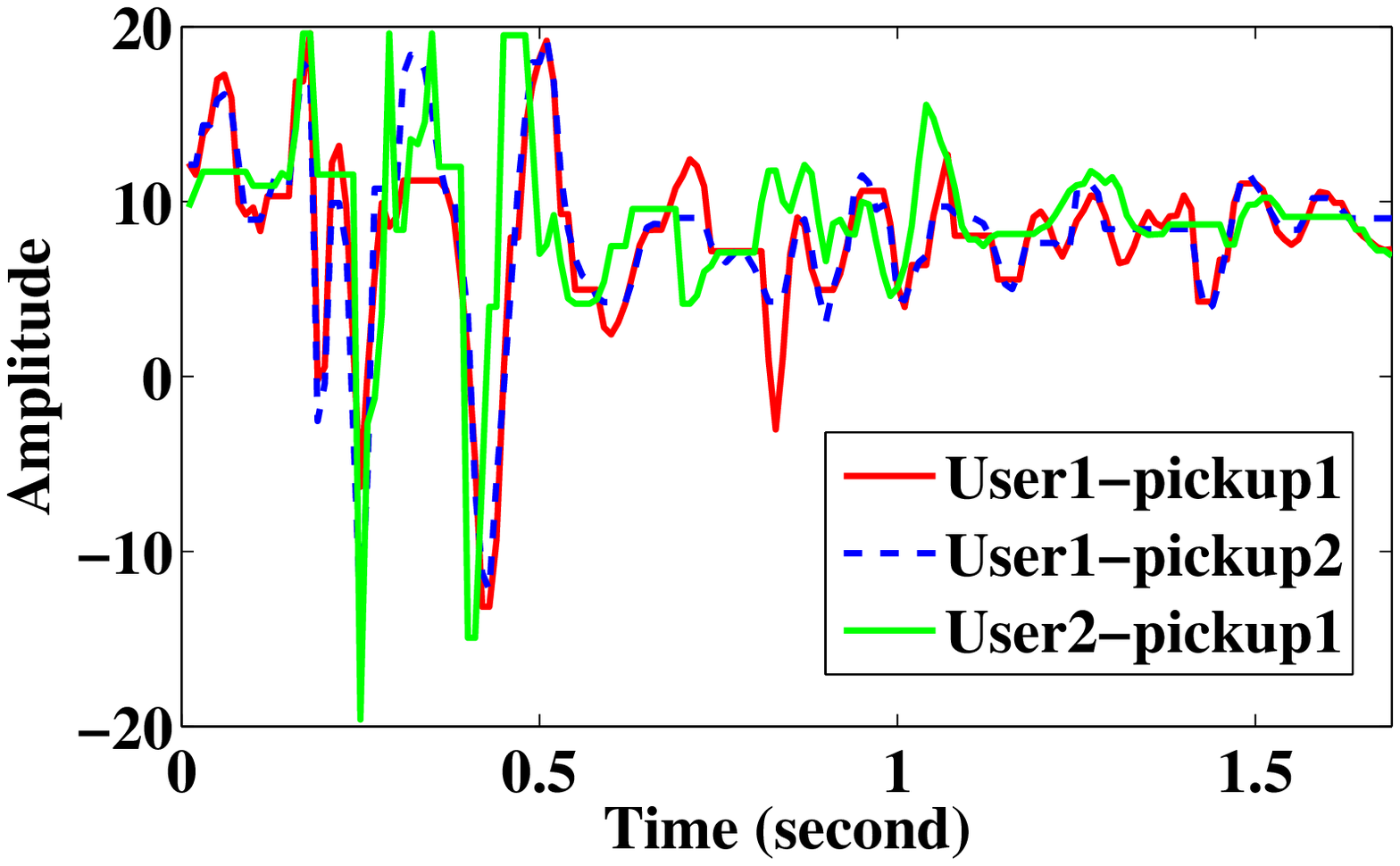, height=1.5 in, width=2.1 in}}
\subfigure[Gyroscope $x$:original signal]{
\label{fig:gyro1_orignal}
\epsfig{file=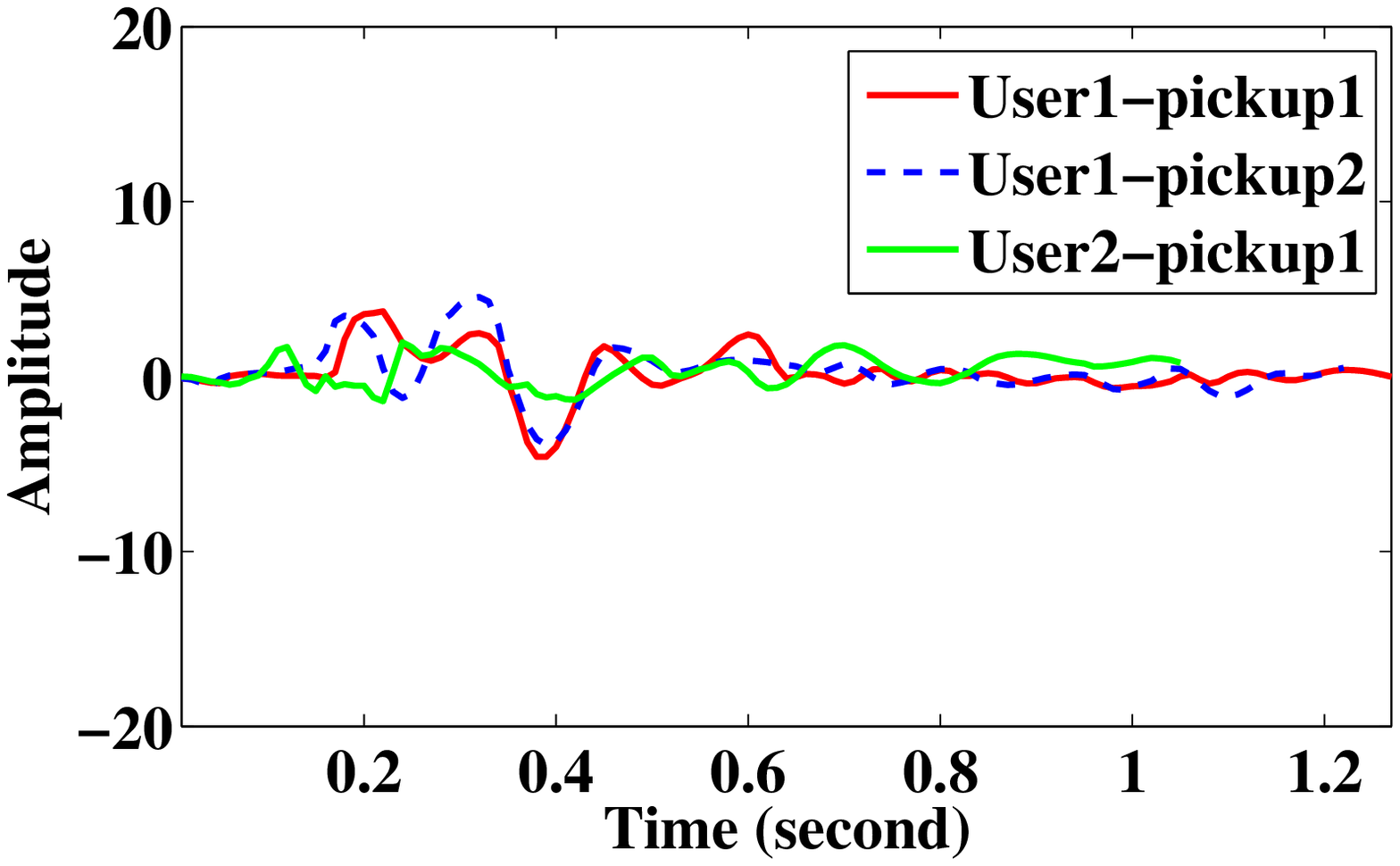, height=1.5 in, width=2.1 in}}
\subfigure[Gyroscope $y$:original signal]{
\label{fig:gyro2_orignal}
\epsfig{file=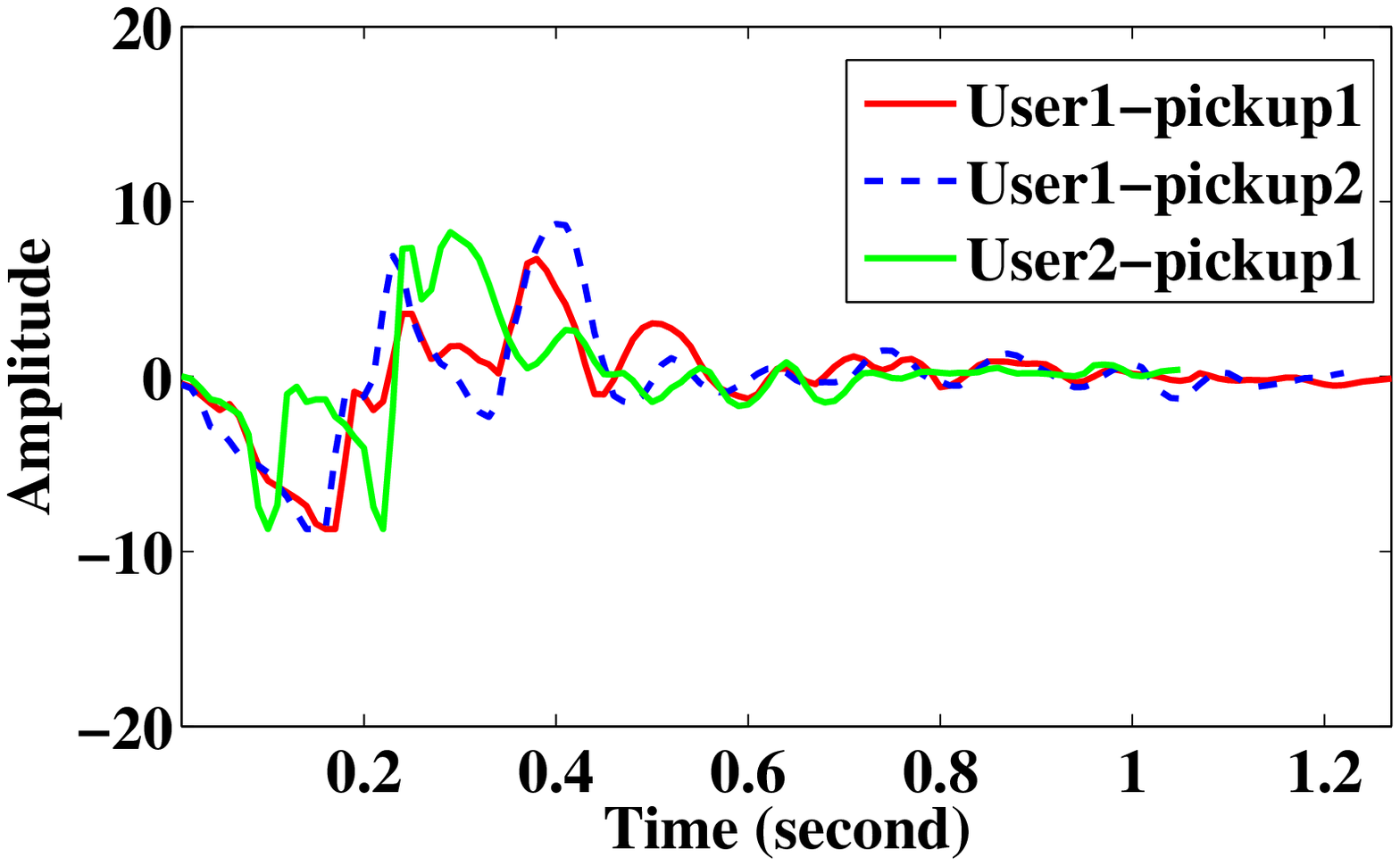, height=1.5 in, width=2.1 in}}
\subfigure[Gyroscope $z$:original signal]{
\label{fig:gyro3_orignal}
\epsfig{file=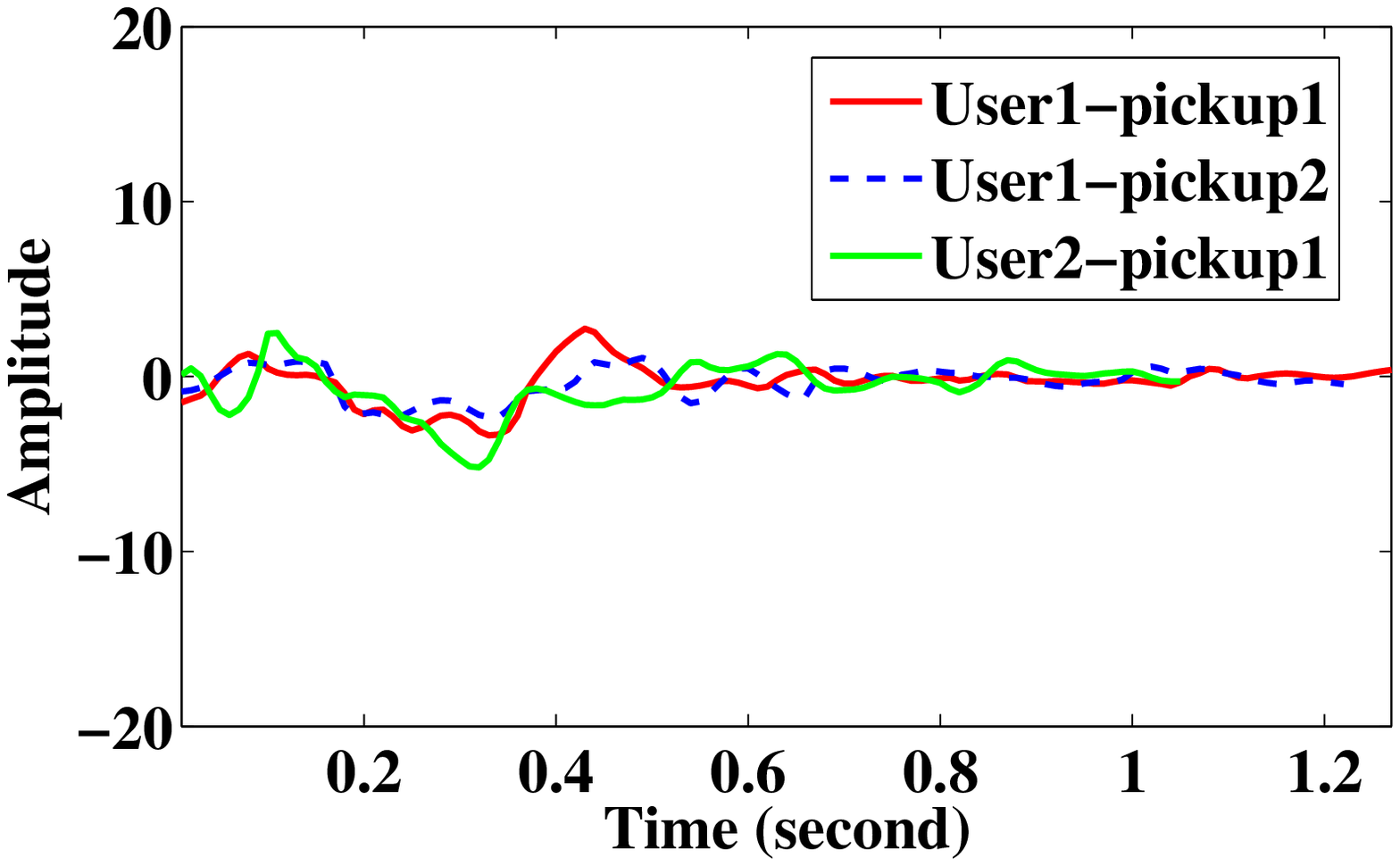, height=1.5 in, width=2.1 in}}
\subfigure[Gyroscope $x$:signal after DTW]{
\label{fig:gyro1_dtw}
\epsfig{file=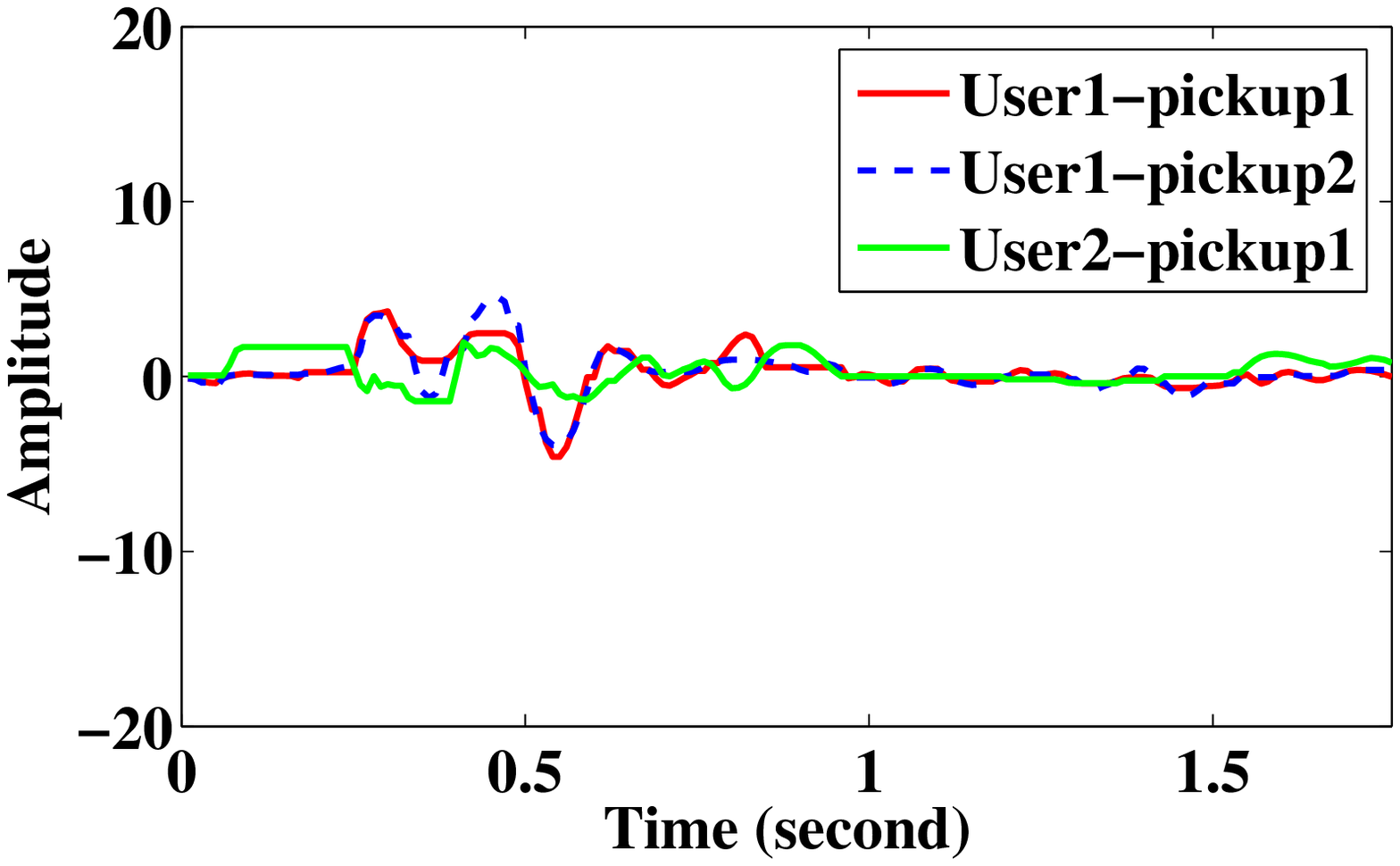, height=1.5 in, width=2.1 in}}
\subfigure[Gyroscope $y$:signal after DTW]{
\label{fig:gyro2_dtw}
\epsfig{file=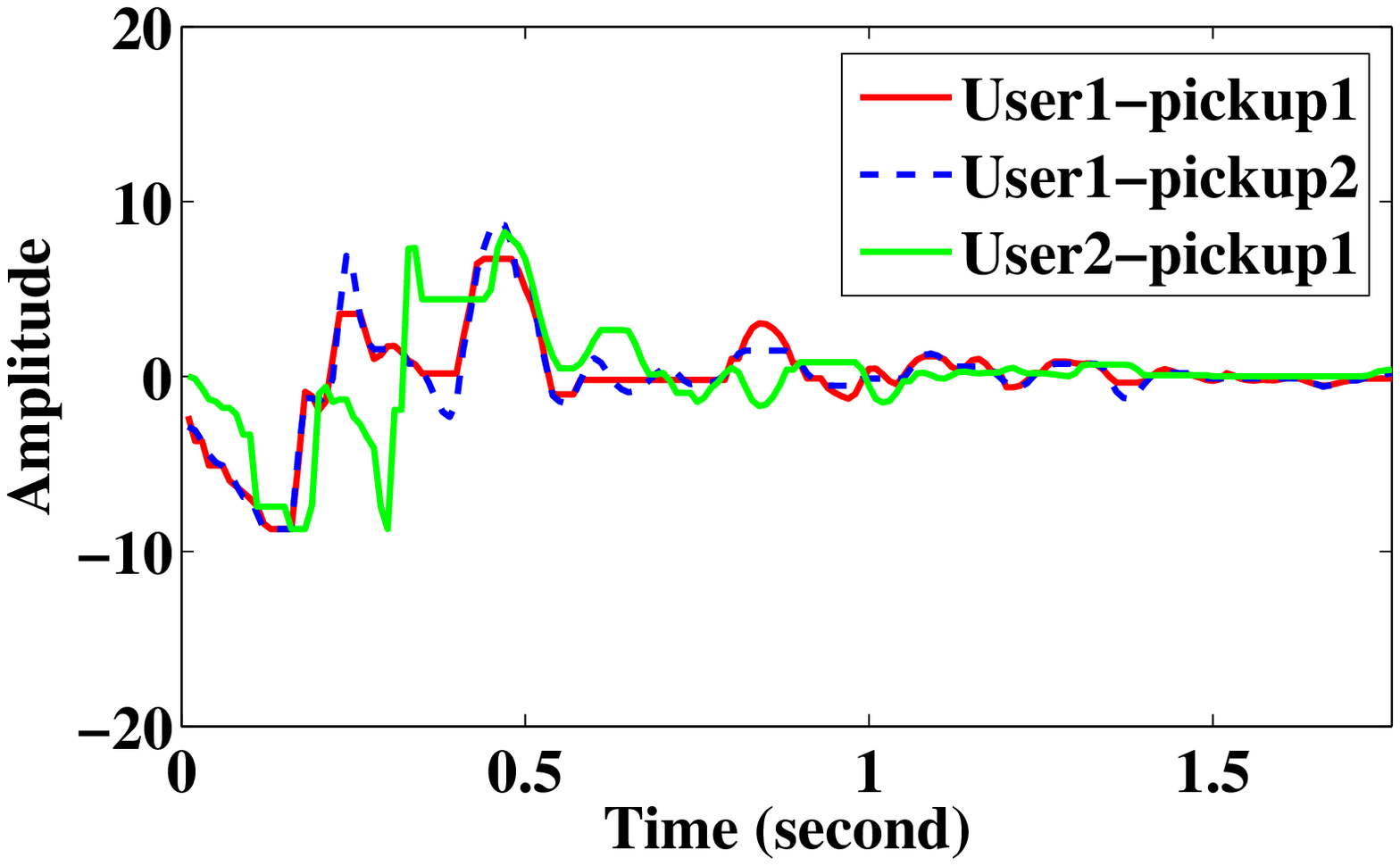, height=1.5 in, width=2.1 in}}
\subfigure[Gyroscope $z$:signal after DTW]{
\label{fig:gyro3_dtw}
\epsfig{file=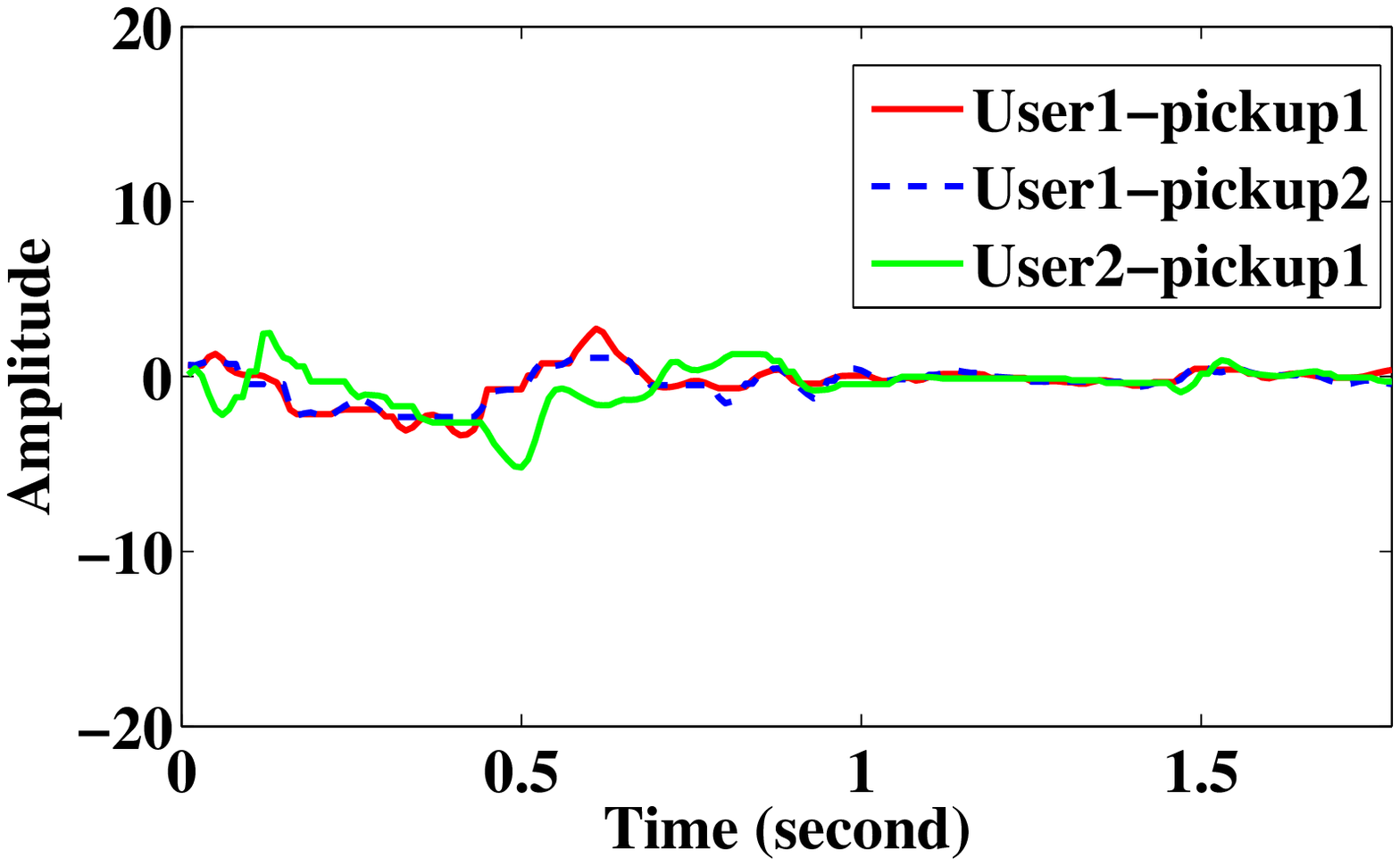, height=1.5 in, width=2.1 in}}
\caption{The visualization of pick-up signals extracted from the accelerometer and gyroscope on three different dimensions. We randomly select two pick-up signals from the same user (red solid and blue dashed dark lines) and a pick-up signal from another user (green light lines). We observe that the distance between two pick-up signals corresponding to the same user is smaller than that from a different user, which lays the foundation for our implicit authentication algorithm. We also observe that different dimensions of sensors may have different powers to distinguish users. For instance, the accelerometer is better than gyroscope in matching the same user's pick-up signals and differentiating different user's patterns, which demonstrates the necessity of our proposed weighted multi-dimensional DTW algorithm.}
\label{fig:visualization}
\end{figure*}

\subsubsection{Multi-dimensional DTW}\label{sec:newDTW}
Different from the popular one-dimensional signal (such as speech signal), each pick-up signal in our setting is multi-dimensional ($6$ dimensions in total including $3$ dimensions for accelerometer and $3$ dimensions for gyroscope), which is a practical challenge for applying the DTW algorithm to our system. In order to address this challenge, we develop a weighted multi-dimensional DTW by carefully analyzing the distinguishing powers of different sensor dimensions.

{\bf Baseline Approach:} \ruby{We} first consider an existing approach to process multi-dimensional signals \cite{shokoohi2015discovery} with DTW as the baseline approach. Consider two $k$-dimensional time-series signals $X:=[X_1,X_2,\dots,X_k]$ and $Y:=[Y_1,Y_2,\dots,Y_k]$, where $X_i$ and $Y_i$ are one dimensional time-series signals for each $i$. Assuming that each dimensional signal is independent of each other, the DTW algorithm under the multiple dimensions setting can be computed as the average over each dimension where

\begin{equation}
DTW_k(X,Y) = \frac{1}{k}\sum_{i=1}^k DTW_1(X_i,Y_i)
\end{equation} 

{\bf Weighted Multi-dimensional DTW:} However, the above baseline approach considers each dimensional signal as contributing equally to the final matching performance, which is an unrealistic assumption. In real world scenarios as in our settings, different dimensions corresponding to different sensors may have varying degrees of influence on the matching performance, since they reflect different levels of a user's behavioral characteristics. Therefore, we propose our weighted multi-dimensional DTW for discriminating the distinguishing powers of different sensor dimensions as:
\begin{equation}\label{eq_weight}
DTW_k(X,Y) = \sum_{i=1}^k w_i DTW_1(X_i,Y_i)
\end{equation} 
where $w_i$ is the weight for the $i$-th dimensional signal.

Figure \ref{fig:visualization} further demonstrates the various distinguishing power for each sensor dimension. We randomly select two pick-up signals corresponding to the same user and one pick-up signal corresponding to another user and compute the distance between these signals after implementing the one-dimensional DTW according to Eq.~\ref{eq_1}. From Figure \ref{fig:visualization},  we observe that the distance between two pick-up signals corresponding to the same user is much smaller than that from a different user, which lays the basic foundation for our implicit authentication algorithm. We also observe that the accelerometer is more powerful than the gyroscope in matching the same user's pick-up signals and differentiating different users' pick-up signals, which demonstrates the empirical necessity of our proposed weighted DTW algorithm. The reason is that a user's pick-up movement is dominated by the translation which is relevant to the accelerometer, while the rotation relevant to the gyroscope is less significant. 

We further analyze the weights for each dimension of accelerometer and gyroscope by varying their weights from $0.1$ to $0.9$ on the axis of $x,y,z$ with summation equal to one. We observe that when each dimension corresponding to the same sensor is equally weighted, the overall authentication performance is the best (with highest authentication accuracy). In addition, we also vary the weights from $0.1$ to $0.9$ on \ruby{the} accelerometer and the gyroscope with summation equal to one. We observe that the best performance (highest authentication  accuracy) is achieved when the ratio between the weight of \ruby{the} accelerometer and that of \ruby{the} gyroscope is $0.6$ to $0.4$. Our observations further demonstrate that the accelerometer is more informative than the gyroscope in improving the authentication performance.

In summary, our SPU system realizes implicit, lightweight and in-device authentication for smartphone users, which consists of sensor data collection, pick-up signal extraction and weighted multi-dimensional DTW processing. If the distance (computed by our multi-dimensional DTW) between two time-series signals is close enough (less than a threshold $\theta$), the user passes the authentication and can have access to the smartphone. 
The detailed process for selecting a proper distance threshold $\theta$ will be described in Section~\ref{threshold}.

\subsection{System Updating}\label{update}
The updating process in previous authentication mechanisms usually involves retraining the authentication classifiers, which is computationally complicated and typically requires additional computing power such as the use of cloud computing. In comparison, we develop an efficient and lightweight updating process to accommodate the user's pick-up behavioral drift over time. 

Our system would automatically update the user's profile in the device whenever the user fails the implicit authentication but successfully passes the subsequent explicit authentication. Our updating process is implemented by averaging the currently stored pick-up profile and the newly-detected pick-up signal. The key challenge for this updating process is that the previous profile and the newly-detected instance may not be of the same length. To solve this problem, we utilize our multi-dimensional DTW algorithm to first scale the two signals to the same length and then average them to obtain the updated user's profile for future authentication. We will show the effectiveness of our system updating process in Section~\ref{second_exp}.

\section{Experiments}\label{sec:experiments}

\begin{figure}[!t]
\centering
\epsfig{file=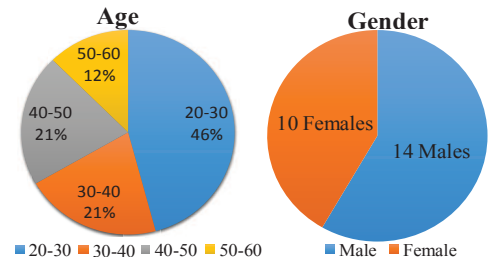, height=1.5 in, width=2.7 in}
\caption{The demographics of users in our experiments.}
\label{fig:demographic}
\end{figure}

\begin{figure*}[!t]
\centering
\subfigure[]{
\label{fig:heatmap_single}
\epsfig{file=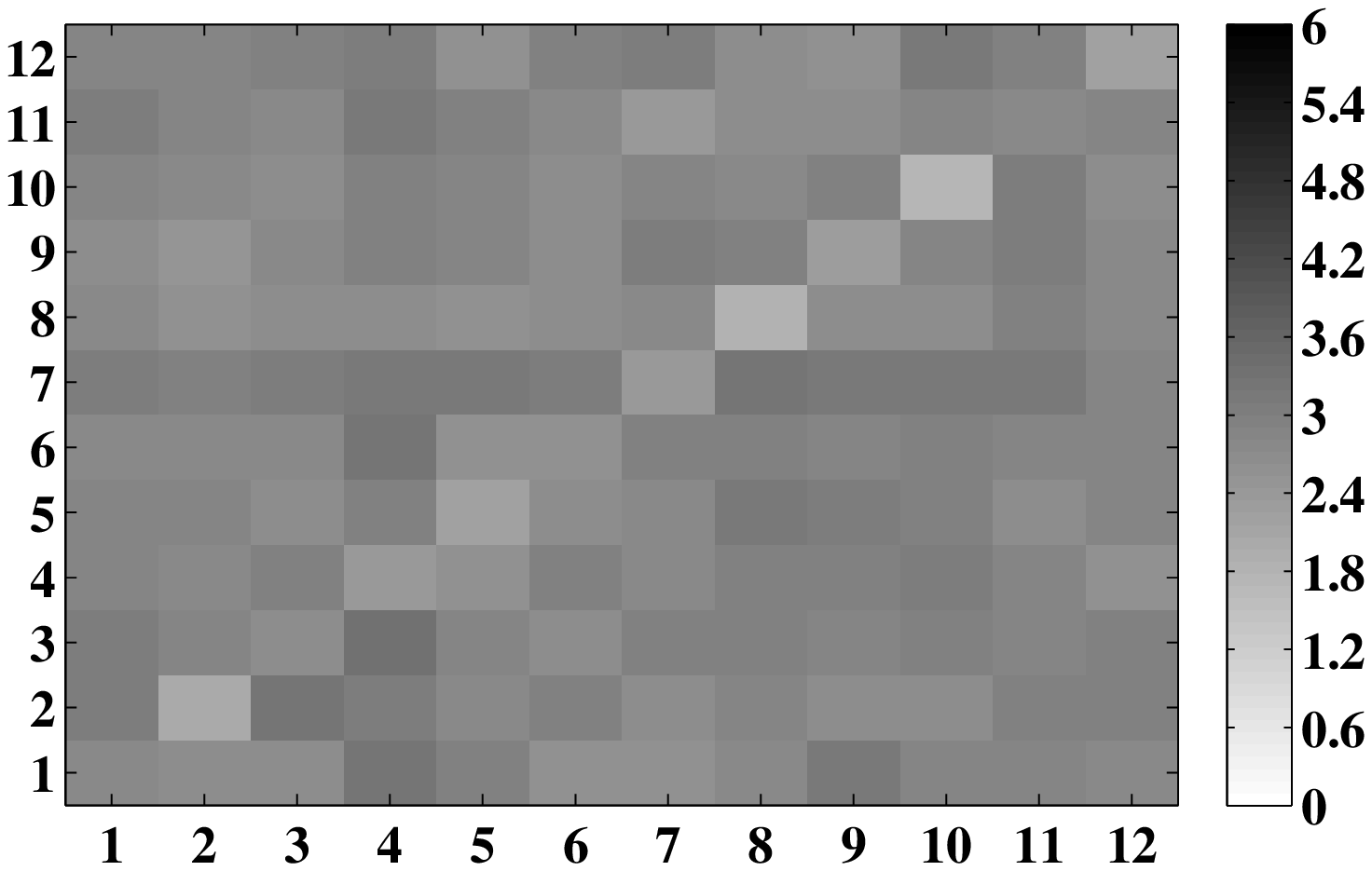, height=1.5 in, width=2.3 in}}
\hspace{0.2in}
\subfigure[]{
\label{fig:heatmap_compare}
\epsfig{file=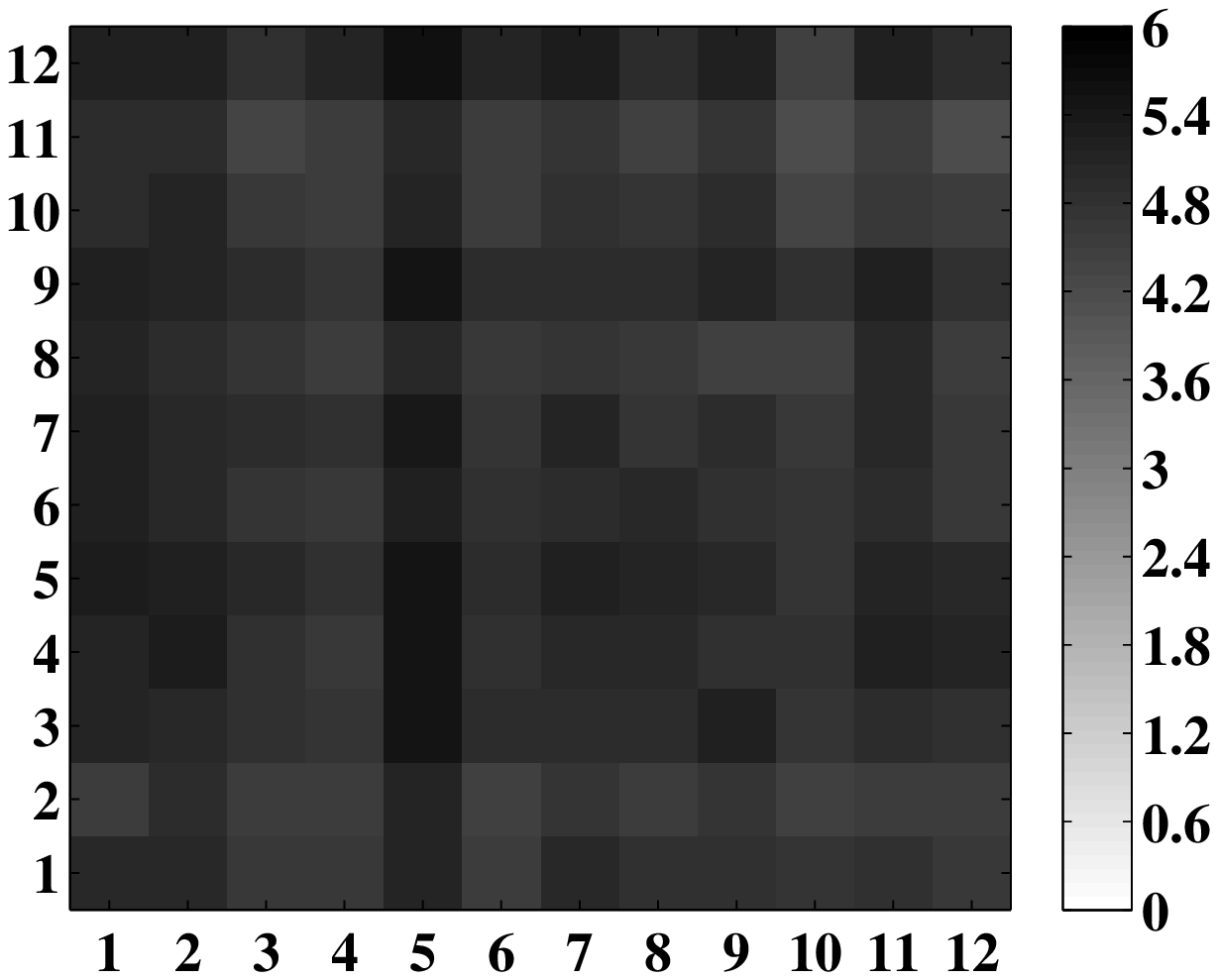, height=1.5 in, width=2.3 in}}
\vspace{-1.5em}
\caption{The heat map of applying weighted multi-dimensional DTW to our dataset. The average DTW distances between different pick-up signals in 12 contexts is collected \ruby{for} all 24 users from (a) the same user, and (b) different users. We can see that the DTW distances from the same user are much lower than that from different users, thus verifying the fundamental intuitions of our proposed algorithm.}
\label{fig:heatmap}
\end{figure*}

To verify the effectiveness of our SPU system, we carefully analyze our collected data (as discussed in Section~\ref{sec:data}) and evaluate the authentication performance of SPU under different experimental scenarios and different system parameters. More specifically, the objectives for our experimental analysis are: 1) to provide empirical confirmation of our system that people's arm movements while they pick up the smartphone can be utilized as a distinguishable behavioral pattern for authentication, as will be discussed in Section~\ref{first_exp}; 2) to investigate the overall authentication performance of SPU under real world usage scenarios,  as will be discussed in Section~\ref{second_exp}; 3) to understand the influence of different system parameters on our system, as will be discussed in Section~\ref{threshold}; 4) to verify the effectiveness of our system updating process (recall Section~\ref{update}), as will be discussed in Section~\ref{incor_update}; 5) to demonstrate the robustness of our system in defending against various impersonation attacks, as will be discussed in Section~\ref{sec:security}; 6) to verify the necessity of combining the accelerometer and gyroscope in our system, as will be discussed in Section~\ref{diff_sensor}.

\subsection{Fundamental Intuition for Our System}\label{first_exp}
Our first experiment was conducted under a lab setting (as described in Section \ref{sec:data}), aiming to demonstrate the fundamental intuition and empirical confirmation for our SPU system. In this experiment, we asked each of the $24$ users to pick up his/her smartphone in $6$ different places while sitting or standing and repeat each movement for $10$ iterations. Figure \ref{fig:demographic} shows the demographics of the $24$ users in our experiments. The average age of the participants is $34.3$ years old while the median is $31$ years old. There are $14$ males and $10$ females.

After extracting the pick-up signals according to Section~\ref{sec:key}, we measure the distance between any two pick-up instances by exploiting the weighted multi-dimensional DTW technique as described in Section \ref{sec:newDTW}. In our algorithm, the weights for the accelerometer signal and the gyroscope signal are selected as $0.6$ and $0.4$ respectively, and each of the $3$ dimensions of the same sensor is weighted equally (recall analysis in Section \ref{sec:newDTW}).

Figure \ref{fig:heatmap_single} shows the average DTW distances of any two instances of pick-up signals corresponding to the same user. Both the $x$-axis and $y$-axis represent the $12$ different pick-up scenarios ($6$ different places and $2$ user states, i.e., sitting or standing). Lighter squares represent smaller DTW distances. In Figure \ref{fig:heatmap_single}, we observe the smallest DTW distances along the diagonal squares since they represent the distances between two pick-up signals corresponding to the same place and user state. By comparing the diagonal squares and the non-diagonal squares in Figure \ref{fig:heatmap_single}, we know the DTW distances across different pick-up scenarios do not vary drastically, demonstrating the robustness of our system under different context scenarios. 

Figure \ref{fig:heatmap_compare} shows the average DTW distances of any two instances of pick-up signals corresponding to different users. From Figure \ref{fig:heatmap}, we observe that the DTW distances between pick-up signals corresponding to the same user are much lower than that between different users, which lays the fundamental intuition for our system that utilizes users' pick-up movements as distinguishable behavioral patterns for authentication.

\subsection{Realistic Usage Scenario}\label{second_exp}
Our second experiment was conducted under a more realistic setting, where the same $24$ users (shown in Figure~\ref{fig:demographic}) were invited to install our SPU application on their own smartphones and use them freely in their normal lives for a week ($7$ days)\footnote{We also let them use our application for another week for evaluating our system updating mechanism as discussed in Section~\ref{incor_update}.}. 

From the collected data, we extracted $3,115$ pick-up signals according to Section~\ref{sec:key}. That is to say, we can detect $18.54$ (i.e., $3115/7/24$) pick-up samples for each user per day (with standard deviation $10.54$). We also recorded the number of times users unlock their smartphones, which is $8,736$ in a week.  Therefore, the average number of times each user unlocks his/her smartphone is $52$ (i.e., $8736/7/24$) per day (with standard deviation $27.31$).  

Note that our system does not detect all the movements when the users try to unlock their smartphones, since we only extract pick-up signals starting from a stable state. In our experiment, we can detect $35.6\%$ (i.e., $18.54/52$, which correspond to the pick-up signals starting from a stable state) of users' pick-up movements when they try to unlock their smartphones. Therefore, we can save more than one third of \ruby{the} time that users need to unlock their smartphones explicitly. Furthermore, we also compute the DTW distance between other types of pick-up signals (e.g., picking up the smartphone from a bag or from a pocket) to investigate whether there are other pick-up patterns  of users that can be utilized for authentication. Our observations show that the distance between other types of pick-up signals (not from a stable state) corresponding to the same user is very large, demonstrating that other types of pick-up signals can not be utilized as distinguishable patterns for user authentication. Therefore, the pick-up movements starting from a stable state which are extracted by our SPU system, constitute the most important pick-up characteristics of users.  Our following experimental analysis are implemented on these detected pick-up movement samples.

\subsubsection{Determining the Distance Threshold}\label{threshold}
A significant challenge in implementing our system is how to select a proper value for the distance threshold $\theta$ between the newly-detected pick-up signal and the stored pick-up profile of the user, which is an important system parameter to balance the trade-off between the usability of our system and the security of smartphone users. A smaller $\theta$ provides higher security, while a larger $\theta$ would result in better usability.

Here, we utilize false acceptance rate (FAR) and false rejection rate (FRR) as metrics to quantify the authentication performance of our system. FAR is the fraction of other users' data that is misclassified as the legitimate user's. FRR is the fraction of the legitimate user's data that is misclassified as other users' data. For security protection, a large FAR is more harmful to the smartphone users than a large FRR. However, a large FRR would degrade the convenience of using our system. Therefore, we aim to investigate the influence of the distance threshold $\theta$ in balancing FAR and FRR, in order to choose a proper $\theta$ for our system.

\begin{figure}[!t]
\centering
\subfigure[]{
\label{fig:threshold}
\epsfig{file=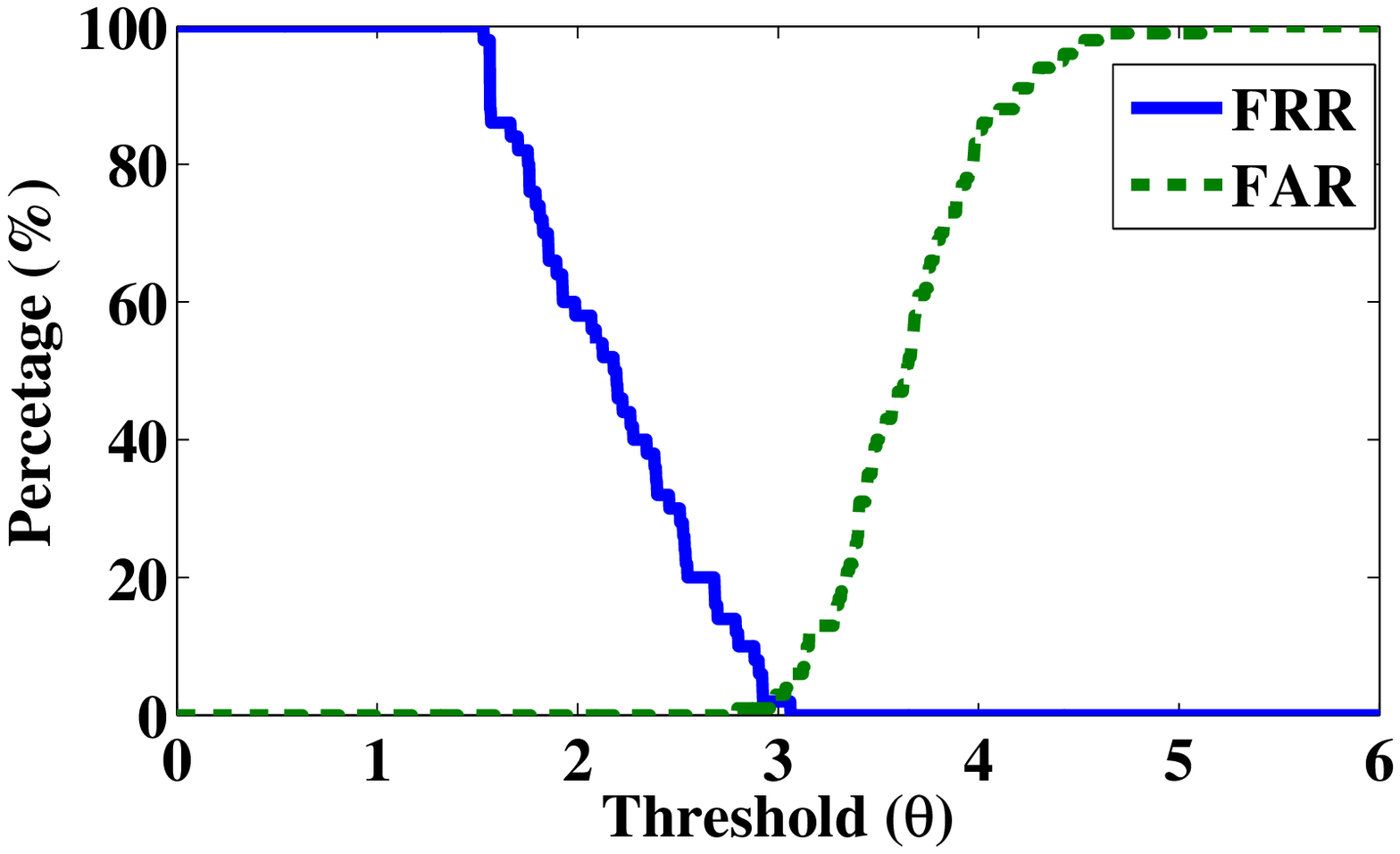, height=1.2 in, width=1.5 in}}
\subfigure[]{
\label{fig:acc_threshold}
\epsfig{file=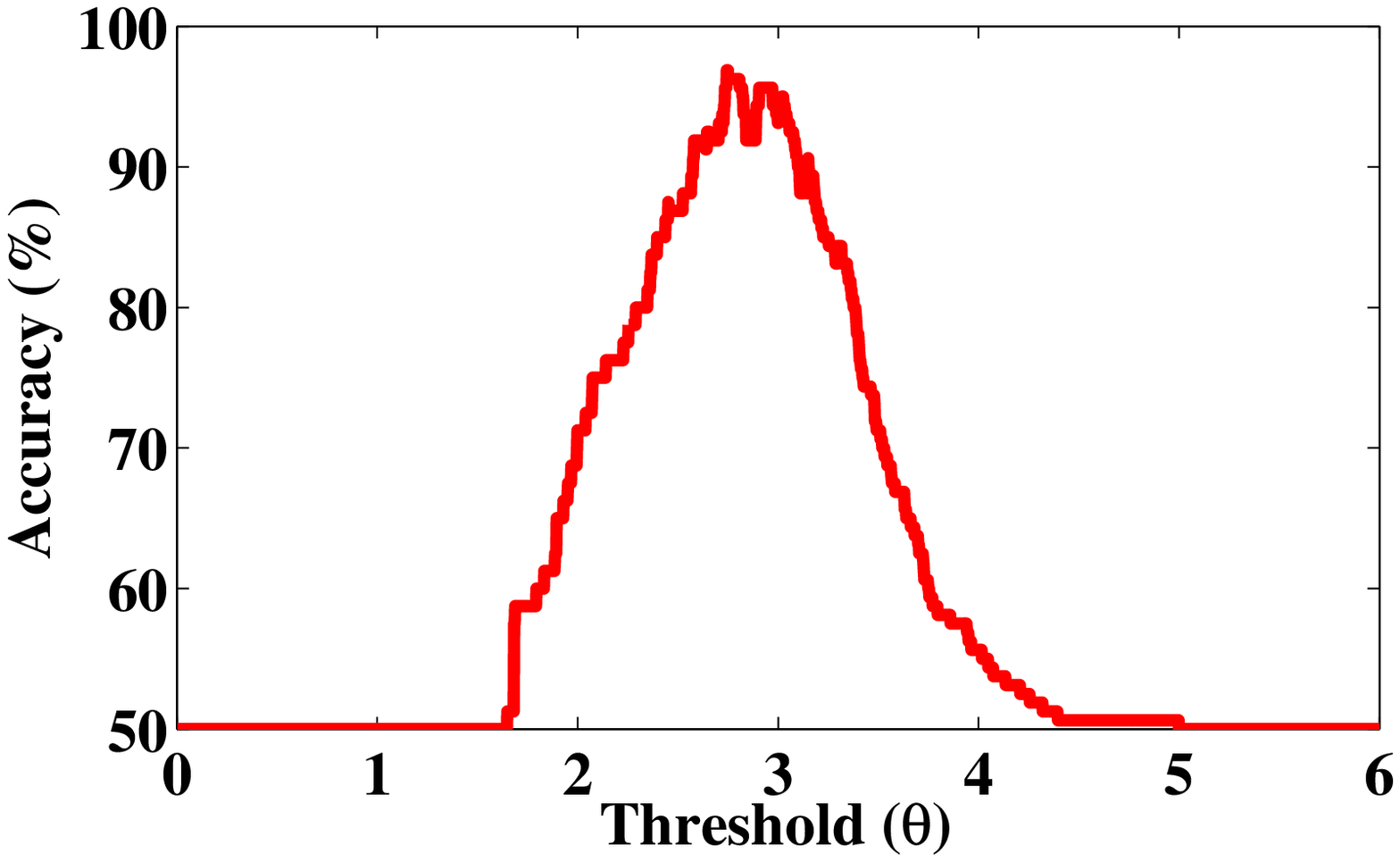, height=1.2 in, width=1.5 in}}
\vspace{-1.5em}
\caption{(a) FAR, FRR and (b) accuracy, varying with different distance threshold $\theta$. We observe that when $\theta = 3.1$, the FRR is $0\%$ and FAR is less than $10\%$. When $\theta = 2.8$ the FRR is $7.6\%$ and FAR is $0\%$, resulting in an authentication accuracy higher than $96.3\%$. Therefore, $\theta$ can tradeoff the usability of our system (lower FRR) and users' security (lower FAR)}
\end{figure}

Figure \ref{fig:threshold} shows the FAR and FRR with varying values of the distance threshold $\theta$. We observe that FAR is less than $10\%$ and FRR is $0\%$ when $\theta = 3.1$. The FAR drops to $0\%$ and FRR increases to $7.6\%$ when $\theta = 2.8$. Therefore, $\theta$ is a trade-off between the usability of our system (lower FRR) and the security of smartphone users (lower FAR). In Figure \ref{fig:acc_threshold}, we observe that the authentication accuracy is higher than $96.3\%$ when $\theta$ is around $2.8$. Combining Figure \ref{fig:threshold} and Figure \ref{fig:acc_threshold}, we choose $\theta=2.8$ in our experiments from now on and in our published system, aiming at minimizing FAR and maximizing the security of the smartphone users.

\subsubsection{Incorporating the System Updating Process}\label{incor_update}
In order to verify the effectiveness of our system updating process as described in Section~\ref{update}, we let the same $24$ users use their smartphones freely for another week. More specifically, we randomly divided the users into two groups. The $12$ users in the first group installed our SPU application which incorporates the updating process, while the other group installed another version of SPU without the updating process. After careful analysis, we observed that the users in the first group needed to explicitly unlock their smartphones (at the same time, their pick-up profiles would be updated in the SPU system) $17$ times per day on average. For the other group without system updating, the users needed to explicitly unlock their smartphones $35$ times per day on average. \hank{We can see that incorporating the system updating process can further reduce $52\%$ of times for users to unlock the smartphones.} These observations show the effectiveness of our system updating process and the advantage of our system in increasing smartphone users' convenience.

\subsection{Security Analysis}\label{sec:security}
In our third experimental setting as described in Section~\ref{sec:data}, we aim to evaluate how robust our SPU system is in defending against various types of impersonation attackers (random attack, context-aware attack and educated attack). 

\begin{figure}[!t]
\centering
\epsfig{file=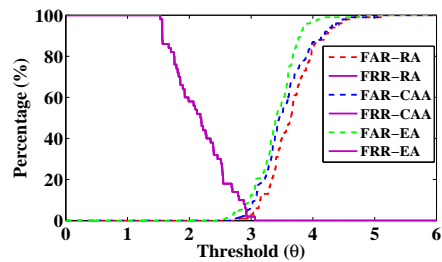, height=1.4 in, width=2.5 in}
\vspace{-1em}
\caption{The FAR and FRR of SPU under various impersonation attacks.}
\label{fig:attacker}
\end{figure}

For each of the three attacks, we computed FAR and FRR curves under different distance thresholds $\theta$ as shown in Figure \ref{fig:attacker}, based on which we have the following observations: 1) SPU can effectively defend against random attacks. Here, `random' attack indicates a brute force attack where the attacker picks up the smartphone randomly without knowing any information about the victim. 2) {When the distance threshold $\theta=2.5$, the FAR becomes $0\%$ for all the three attacks and the corresponding FRR is $18\%$.} Note that the FRR curve for the three attacks are the same since it evaluates the ratio that the victim is rejected by our system, which is irrelevant to the attacker's capability. 3) Furthermore, the user can defend against different levels of attacks by adjusting the distance threshold $\theta$. These results suggest that our SPU system is more robust against random (brute force) attacks than other types of impersonation attacks (context-aware attacks and educated attacks) since these advanced attackers usually have access to partial information about the user's pick-up movements (recall Section~\ref{threat} and Section~\ref{sec:data}).

In summary, SPU can defend against most realistic attacks robustly and effectively. Even with a strong attacker (i.e., an insider attacker), our system performs gracefully.

\subsection{Further Experiments}\label{diff_sensor}
We further demonstrate the necessity and advantages of combining the common sensors, acclerometer and gyroscope, in our SPU system. In Table \ref{table:sensor}, we observe that using the combination of accelerometer and gyroscope can achieve better performance than using each sensor individually, with the authentication accuracy up to $96.3\%$. Furthermore, our SPU can reduce the number of explicit authentications a user must do by $32.9\%$ (i.e., $35.6\% \times (1-7.6\%)$) on average, where $35.6\%$ is the ratio of detected pick-up signals (recall Section~\ref{second_exp}) and $7.6\%$ is the FRR by using the combination of accelerometer and gyroscope.

Next, we went a step further to investigate whether our SPU system could benefit from more sensors than just the accelerometer and gyroscope. More specifically, we analyze the authentication performance of SPU when incorporating the measurements of a magnetometer and its combinations with the accelerometer and gyroscope. We consider the magnetometer since we can construct the popular $9$-axis motion detector of the smartphone by combining the $3$-axis measurements of magnetometer with the $3$-axis measurements of each of accelerometer and gyroscope. An interesting observation shown in Table \ref{table:sensor_magnetic} is that incorporating the magnetometer into our SPU system does not improve the overall authentication accuracy - in fact, it degrades the authentication accuracy! Using more sensors is not always better! The reason is that the magnetic field is rather sensitive to the direction of the smartphone, which makes it vary significantly when the same user picks up the smartphone in different directions - thus degrading the overall authentication performance. 

These observations substantiate our choice of using only the accelerometer and gyroscope in our system.

\begin{table}[!t]
\small
\centering
\caption{
\small
The authentication accuracy by using accelerometer and gyroscope with distance threshold $\theta=2.8$.}
\begin{tabular}{|c|c|c|c|} \hline
   & Accuracy & FAR & FRR \\ \hline
Accelerometer & 90.9 \% & 6.4\% & 11.8\% \\ \hline
Gyroscope 		& 85.2 \% & 13.7\% & 15.2\% \\ \hline
Acc+Gyr  & 96.3 \% & 0\% & 7.6\% \\ \hline
\end{tabular}
\label{table:sensor}
\end{table} 

\begin{table}[!t]
\small
\centering
\caption{
\small
The authentication accuracy by using three motion sensors with distance threshold $\theta=2.8$.}
\begin{tabular}{|c|c|c|c|} \hline
   & Accuracy & FAR & FRR \\ \hline
Magnetometer & $36.7 \%$ & $54.4\%$ & $62.4\%$ \\ \hline
Acc+Mag  & $67.2 \%$ & $37.2\%$ & $48.7\%$ \\ \hline
Gyr+Mag  & $54.8 \%$ & $41.9\%$ & $57.1\%$ \\ \hline
All three sensors & $72.5 \%$ & $27.6\%$ & $34.4\%$ \\ \hline
\end{tabular}
\label{table:sensor_magnetic}
\end{table} 

\section{Overhead Analysis}\label{sec:overhead}

We now evaluate the system overhead of SPU on personal smartphones to demonstrate the applicability of our system in real world scenarios. In our source code, the DTW algorithm is implemented in the C language by using the Native Development Kit (NDK) in Android 5.1. We test our system on a Google Nexus5 with $2.3$GHz, Krait $400$ processor, $16$GB internal storage and $2$GB RAM, using Android 5.1. 
\subsection{Power Consumption}
There are four different testing scenarios: 1) Phone is locked and SPU is off; 2) Phone is locked and SPU keeps running; 3) Phone is under use and SPU is off; 4) Phone is under use and SPU is running. 

For cases 1) and 2), the test time is 12 hours each. We charge the smartphone battery to 100\% and check the battery level after 12 hours. The average difference of battery charged level from 100\% is reported in Table \ref{table:power}. For cases 3) and 4), \emph{the phone under use} means that the user keeps unlocking and locking the phone. During the unlocked time, the user keeps typing notes. The period of unlocking and locking is two minutes and the test time in total is 60 minutes. 

Table \ref{table:power} shows the result of our power consumption test on battery usage. We find that in cases 1) and 2), the SPU-on mode consumes 1.8\% more battery power than the SPU-off mode each hour. We believe the extra cost in battery consumption caused by SPU will not affect user experience in daily use. For cases 3) and 4), SPU consumes 2\% more battery power performing 30 SPU implicit authentications in one hour, which is also an acceptable cost for daily usage. 
\begin{table}[!t]
\small
\centering
\caption{
\small
The power consumption under four different scenarios.}
\begin{tabular}{|l|c|} \hline
Scenario & \tabincell{c}{Power \\Consumption} \\ \hline
1) Phone locked, SPU off & $1.1 \%$ \\ \hline
2) Phone locked, SPU on & $2.9 \%$ \\ \hline
3) Phone unlocked periodically, SPU off & $1.5 \%$ \\ \hline
4) Phone unlocked periodically, SPU on & $3.5 \%$ \\ \hline
\end{tabular}
\label{table:power}
\end{table} 
\subsection{Response Time}
\begin{figure}
\centering
\epsfig{file=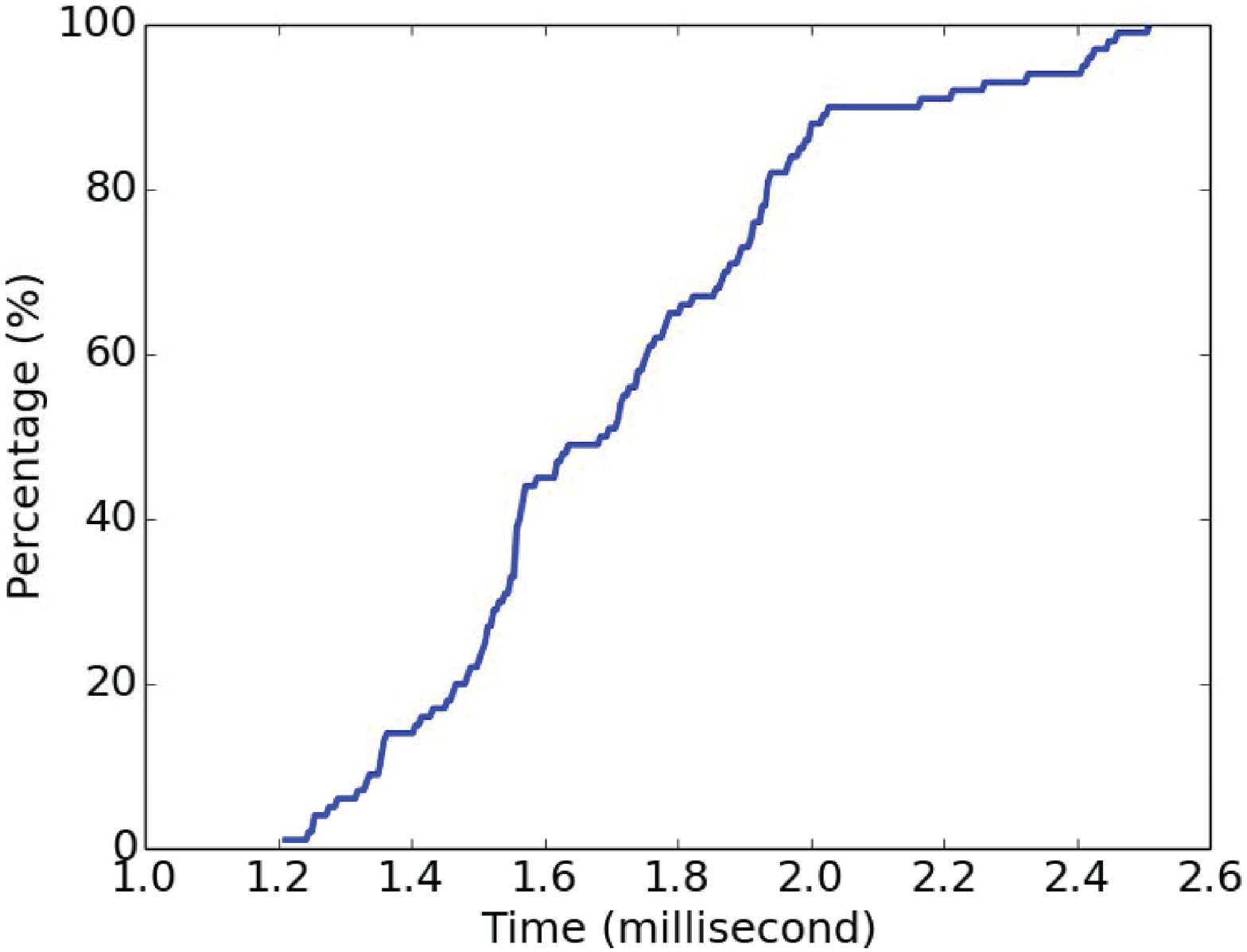, height=1.3 in, width=2.2 in}
\vspace{-1em}
\caption{Cumulative distribution function of decision-making time in SPU. We can find that more than $90\%$ of decision-making processes can be completed within $2$ milliseconds and all the processes can be finished within $2.4$ milliseconds.}
\label{fig:latency}
\end{figure}

Figure \ref{fig:latency} shows the cumulative distribution function of decision-making time in SPU authentication. We find that more than $90\%$ of the decision-making computations can be completed within $2$ milliseconds and all can be finished within $2.4$ milliseconds. This result shows that the latency caused by the SPU system for authentication is low enough to be user-friendly and reasonable for normal usage.

\section{Related Work}\label{sec:related}

User authentication is one of the most important issues in smartphone security. Password-based authentication approaches are based on possession of secret information, such as passwords or PINs. Biometric-based approaches make use of distinct personal features, such as fingerprint or iris patterns. Behavior-based authentication identifies a user based on his/her behavioral pattern that is observed by the smartphone. Compared with the password-based and the biometric-based authentication, the behavior-based authentication is more convenient for smartphone users with good resilience to forgery attacks.

\subsection{Password-based Authentication}
The objective of most password-based authentication mechanisms, e.g., PIN or passwords, is to secure the phone from unwanted access. However, these methods require frequent participation of the user. This often leads to interruptions to the smartphone user, e.g. continuously prompting him/her with some challenges. As a result, many smartphone users tend to completely remove such authentication methods \cite{report_convenience}. \
Our SPU system can overcome these weaknesses, which increases the convenience for smartphone users while guaranteeing their security, as shown in Section \ref{sec:experiments}.
\subsection{Biometric-based Authentication}
Biometric-based authentications study static physical features of humans. Currently, there are many different physiological biometrics for authentication, such as face patterns, fingerprints \cite{hong1998pami}, and iris patterns \cite{qi2008iris}. Biometric-based authentication systems involve an enrollment phase and an authentication phase. A user is enrolled by providing his/her biological data such as fingerprint or iris pattern. The system extracts these patterns from the provided data and stores the extracted patterns for future reference. During the authentication phase, the system compares the observed biological data against the stored data to authenticate a user.

However, biometric-based authentications also require frequent user participation, and hence is also an explicit authentication mechanism. For example, fingerprint authentication always requires the user to put his/her finger on the fingerprint scanner. On average, the response time is longer than $1$ second \cite{report_fingerprint}, which is also much longer than the $2.4$ milliseconds of our SPU system. Hence, unlike our implicit SPU authentication, these biometric-based approaches requiring user compliance are not as convenient as our SPU system.
\subsection{Behavior-based Authentication}
Another thread of authentication research measures the behavioral patterns of the user, where a user is identified based on his/her behavioral patterns, such as hand-writing pattern \cite{cc5,frank2013touchalytics}, gait \cite{cc7} and GPS location patterns \cite{cc4}. 

With the increasing development of mobile sensing technology, collecting measurements through sensors built within the smartphone and other devices is now becoming not only possible, but quite easy through, for example, Android sensor APIs. Mobile sensing applications, such as the CMU MobiSens\cite{cc1}, run as services in the background and can constantly collect sensors' data from smartphones. Sensors can be either hard sensors (e.g., accelerometers) that are physically-sensing devices, or soft sensors that record information of a phone's running status (e.g., screen on/off). Therefore, sensor-based implicit authentication mechanisms have become very popular and applicable for behavior-based authentication.

In \cite{cc4}, an $n$-gram geo-based model is proposed for modeling a user's mobility pattern. They use the GPS sensor to detect abnormal activities (e.g., a phone being stolen) by analyzing a user's location history, and their algorithm can achieve $86.6\%$ accuracy. However, the access to GPS require users' permissions, and cannot be done implicitly.

Nickel et al.~\cite{cc7} exploited a user's walking pattern to authenticate a smartphone user by using the $k$-NN algorithm.
Conti et al.~\cite{conti2011swing} utilized the user's movement of answering a phone call to authenticate a smartphone user. \hank{Shrestha et al.\cite{shrestha2016theft} utilized a tapping pattern to authenticate a user when the user \ruby{does an} NFC transaction.}
However, their experiments had strict restrictions on the users' behavior where the users have to walk or answer a phone call following a specific script (e.g., walk straight ahead at the same speed~\cite{cc7} or answer the phone which is on a table in front of a user~\cite{conti2011swing}). These restrictions are impractical for \ruby{a} real use.

Users' behavior on \ruby{a} touch screen is one of the most popular research directions in behavior-based authentication \cite{cc5, frank2013touchalytics, cc6, sherman2014user, buschek2016evaluating}.
Trojahn et al. \cite{cc5} developed a mixture of a keystroke-based and a handwriting-based method to realize authentication by using the screen sensor. Their approach has achieved $11\%$ FAR and $16\%$ FRR.
Frank et al. \cite{frank2013touchalytics} studied the correlation between $22$ analytic features from touchscreen traces and classified these features using $k$-NN and SVM.
Li et al. \cite{cc6} proposed another behavior-based authentication method where they exploited five basic movements (sliding up, down, right, left and tapping) and their related combinations, as the user's behavioral pattern features, to perform authentication.
However, touch screen based authentications may suffer from a simple robotic attack \cite{serwadda2013kids}.

SenSec~\cite{cc2} constantly collects data from the accelerometer, gyroscope and magnetometer, to construct gesture models while the user is using the device. SenSec has shown that it can achieve $75\%$ accuracy in identifying owners and $71.3\%$ accuracy in detecting the adversaries.
Lee et al.~\cite{lee2015icissp} monitored the users' general behavioral patterns and utilized SVM techniques for user authentication. Their results show that the authentication accuracy can be higher than $90\%$ by using a combination of sensors. However, these methods require a large amount of privacy sensitive training data from other users, and significant external computation power for learning the behavior models, unlike our in-device SPU authentication method. 

In fact, almost all the existing behavior-based authentication mechanisms~\cite{cc5,cc7,cc2,cc4,cc6,lee2015icissp,frank2013touchalytics} 
heavily rely on a powerful remote server to share the tasks and take a relatively long time to complete the authentication process. In comparison, our SPU is a lightweight, in-device, non-intrusive and automatic-learning authentication system, which would increase the convenience for smartphone users while enhancing their security.

\section{Discussion and Future Work}

Our SPU system increases the convenience for smartphone users while enhancing their security. We will make SPU open source software, suitable for extensions with future research and experiments.

Future research can include more context-detection techniques to detect fine-grained pick-up patterns for users and embed it with SPU to further increase the convenience and security for smartphone users.

Users' pick-up patterns may vary when they are using other types of devices, e.g., tablets or smartwatches. It would be an interesting future direction to extend SPU to these mobile devices. Furthermore, the combination of multiple devices may possibly provide better authentication performance for the SPU system.

\section{Conclusion}

We proposed a novel system, Secure Pick Up (SPU), to implicitly authenticate smartphone users in a lightweight, in-device, non-intrusive and automatic-learning manner. Unlike previous work, SPU does not require a large amount of training data (especially those of other users) or any additional computational power from a remote server, which makes it more deployable and desirable for many users.

Our key insight is that the user's phone pick-up pattern is distinguishable from others, using smartphone sensor measurements. We propose a weighted multi-dimensional dynamic time warping algorithm to effectively measure the distance between pick-up signals in order to determine the legitimate user versus others. 

Extensive experimental analysis shows that our system achieves authentication accuracy up to $96.3\%$ with negligible system overhead ($2\%$ power consumption). Furthermore, our evaluation shows that SPU can reduce by $32.9\%$ the number of explicit authentications a user must do, and can defend against various impersonation attacks effectively. Overall, SPU offers a novel feature in the design of today's smartphone authentication and provides users with more options in balancing the security and convenience of their devices.

\bibliographystyle{ACM-Reference-Format}
\bibliography{sigproc} 

\end{document}